\documentclass[12pt]{article}
\usepackage{amsmath}
\usepackage{graphicx}
\usepackage{enumerate}
\usepackage[numbers]{natbib}
\usepackage{url} 
\usepackage{amsthm,amsmath,amsfonts,amssymb}

\newcommand{\blind}{1}

\usepackage{amsthm,amsmath,amsfonts,amssymb}
\RequirePackage[colorlinks,citecolor=blue,urlcolor=blue]{hyperref}
\RequirePackage{graphicx,xcolor}
\usepackage{enumerate}
\usepackage{multirow}
\usepackage{bm}
\usepackage{hyperref}
\usepackage[all]{xy}
\usepackage{graphicx,caption, subcaption}
\usepackage{float}
\usepackage{color}
\usepackage{booktabs}
\usepackage{array}
\newcolumntype{P}[1]{>{\raggedright\arraybackslash}p{#1}}

\usepackage{enumitem}
\usepackage{makecell}

\numberwithin{equation}{section}

\newtheorem{theorem}{Theorem}[section]

\theoremstyle{remark}
\newtheorem{definition}[theorem]{Definition}

\newcommand{\bed}{\begin{definition}}
	\newcommand{\eed}{\end{definition}}

\newcommand{\rom}[1]{\uppercase\expandafter{\romannumeral #1\relax}}

\usepackage{bbm}

\newcommand{\bitem}{\begin{itemize}}
	\newcommand{\eitem}{\end{itemize}}

\newcommand{\goto}{\rightarrow}

\newcommand{\beqn}{\begin{equation}}
	\newcommand{\eeqn}{\end{equation}}
\newcommand{\balign}{\begin{align}}
	\newcommand{\ealign}{\end{align}}

\usepackage{amssymb}
\newcommand{\beq}{\begin{equation}}
	\newcommand{\eeq}{\end{equation}}

\newcommand{\diag}{\mathrm{diag}}

\allowdisplaybreaks

\usepackage{enumitem}
\usepackage{makecell}

\usepackage{listings}
\begin{document}

\def\spacingset#1{\renewcommand{\baselinestretch}%
{#1}\small\normalsize} \spacingset{1}


\if1\blind
{
  \title{\bf Recent Advances in Text Analysis}
  \author{     Zheng Tracy Ke, 
  Pengsheng Ji, 
    Jiashun Jin, and Wanshan Li\\
    Harvard University, University of Georgia, and  Carnegie Mellon University}
  \maketitle
} \fi

\if0\blind
{
  \bigskip
  \bigskip
  \bigskip
  \begin{center}
    {\LARGE\bf Supplement of ``Recent Advances in Text Analysis"}
\end{center}
  \medskip
} \fi



%
%
%

\begin{abstract}
Text analysis is an interesting research area in data science and has various applications, such as in artificial intelligence, biomedical research, and engineering. We review popular methods for text analysis, ranging from  
topic modeling to the recent neural language models. 
In particular, we review Topic-SCORE, a statistical approach to topic modeling, and discuss how to use it to analyze MADStat - a dataset on statistical publications 
that we collected and cleaned.

The application of Topic-SCORE and other methods on MADStat leads to interesting findings.  
For example,  
$11$ representative topics in statistics are identified.  
For each journal, the evolution of 
topic weights over time can be visualized, and these results are used to analyze the trends in 
statistical research. In particular,  we propose a new statistical model for ranking the citation impacts of $11$ topics, and we also 
build a cross-topic citation graph to illustrate how research results on different topics spread to one another.

The results on MADStat provide a data-driven picture of the statistical 
research in $1975$--$2015$, from a text analysis perspective.  
\end{abstract}


\bigskip

\noindent
{\bf Data and code:}
The data and code for text analysis conducted in this article can be found at multiple repositories, including the journal website (\url{https://www.annualreviews.org/doi/abs/10.1146/annurev-statistics-040522-022138}), GitHub (\url{https://github.com/ZhengTracyKe/MADStat-Text}), and Harvard Dataverse (\url{https://dataverse.harvard.edu/dataset.xhtml?persistentId=doi:10.7910/DVN/YIXS6B}).


\newpage
\spacingset{1.475} 

\tableofcontents

\section{Introduction}  
\label{sec:intro}  
Text analysis is an interdisciplinary 
research area in data science, computer science, and 
linguistics. It aims to use computers to process a large amount of 
natural language data and extract information or features.  
Research in text analysis and Natural Language Processing (NLP) 
is especially useful for developing auto-piloting cars, chatbots (e.g., chatGPT),  
and artificial intelligence in health care and biomedical engineering.  
In the past decades, numerous methods were proposed for text analysis. 
Two approaches are especially popular. 
\begin{itemize} 
\setlength{\itemsep}{-0.015 em} 
\item {\it Topic modeling}. This approach has a strong statistical flavor.  
Given a large collection of text documents, this approach assumes that all these documents only discuss a few topics (e.g., ``finance", ``politics", ``sports", etc.). 
Each document discusses the topics with 
different weights, and given that a particular topic is being discussed,  the words in the document are generated from a distribution 
specific to that  topic. 
\item {\it Neural network modeling}. This is a rapidly developing area. It models the generation of text documents via  
deep neural networks, and train the model with massive text corpora (e.g., English Wikipedia) and domain knowledge.  The trained model will be used for different down-stream tasks. 
\end{itemize} 
The neural network approach has proven effective 
in many NLP tasks (e.g., text classification and machine translation), 
and has gained immense popularity, particularly among technology titans such as Google and Meta. However,  this approach is internally complex, expensive to train, and resource-intensive. 
These factors substantially restrict the use of the neural network approach, especially for some common NLP users such as social scientists who only have a few hundreds of text documents from a specific domain of interest. The topic modeling approach provides a valuable alternative and has the following benefits.
\begin{itemize} 
\item ({\it Transparency and interpretability}). Many users prefer an approach that is (a) {\it not a blackbox} but a more transparent step-by-step algorithm,  (b) easy to understand and tune (so users can modify it as needed), and (c) 
where the results (e.g., the extracted features)  are easy-to-interpret (see \cite{donoho201750,donoho2015higher}). 
\item ({\it Analytical accessibility}).  Topic modeling approaches 
are relatively simple and allow  for  delicate theoretical analysis. Especially, some of these methods are shown to enjoy statistical optimality.  In comparison, neural network approaches are much harder to analyze and often have no theoretical guarantee. 
\end{itemize}  
Topic-SCORE \citep{KaW2017} is an especially interesting topic modeling method. 
It is fast, effective, and enjoys nice theoretical properties. 
It is also a flexible idea and can adapt to several different 
settings.  These characteristics make Topic-SCORE especially appealing when we analyze the 
  MADStat data set (to be introduced below).  

One goal of this paper is to review popular topic modeling methods, 
from the 
rudimentary topic models in the 1990s to the more recent multi-gram 
topic models, 
with a focus on Topic-SCORE and related problems. 
In addition, we review the neural network approaches. 
Large neural language model is a rapidly developing area (with new 
research emerging on a weekly basis), making it hard to conduct a comprehensive review. 
Since the focus of this paper is on the topic modeling approach and the MADStat data set, we keep the review of neural network approaches relatively brief.

Another goal of this paper is to analyze the MADStat dataset using 
text analysis techniques.  MADStat \citep{ji2022co} is a large-scale high-quality data set on statistical publications.    We collected and cleaned the dataset,  with substantial time and efforts.  
It consists of the bibtex (title, author, abstract, keywords, references) and citation information  of 83,331 research papers published in 36 representative journals in statistics and related fields during 1975--2015. 
The dataset contains {\it detailed citation, bibtex, and author information for each paper}  
(aka. paper-level data). It can be used to study research problems that can not be addressed with other data resources that have only journal-level data or include no author information.  Using MADStat, for instance, one can easily find the top 30 most-cited papers within our data range, whereas it is unclear how to do so using Google Scholar.

Text analysis on MADStat yields several findings. First,  
we use Topic-SCORE to identify $11$ representative research topics in statistics, and 
visualize the evolution of the overall weight of statistical publications on each topic. 
Second, we extend Topic-SCORE to TR-SCORE, a method for ranking 
research topics by their citation exchanges, and we also build a knowledge graph to visualize how 
the research results on one topic disseminate to others.  
Third, we rank all $36$ journals and suggest that Annals of Statistics, Biometrika,  JASA, and JRSS-B 
are the four most influential journals in statistics. 
Last, we find that the (per-author) paper counts in statistics were steadily decreasing, 
suggesting that publishing in statistics has becoming more and more competitive. 
Our results provide an evidence-based picture of the whole statistics community,
and so can be viewed as a data-driven review of statistical research, 
from a text analysis persective. 
The results may help administrators or committees for decision making (e.g., promotion and award) and
help researchers make research plan and build networks. We use statistics as the object of study, but the same techniques can be used to study other fields (e.g., physics).


Obtaining a large-scale, high-quality data set such as MADStat is a challenging and time-consuming task. Particularly, many public data (e.g., Google Scholar) are quite noisy, and many online resources do not permit large-volume downloads. 
The data set must also be carefully cleansed; we accomplish this through a combination of manual labor and custom-developed computer algorithms. See Section~\ref{sec:data-collection} of the supplement for more detailed discussion on data collection and cleaning.  

Below in Section \ref{sec:topic}, we review the recent advances on topic modeling. In  Section \ref{sec:DNN}, we briefly review  
neural network language models. 
In Section \ref{sec:basic}, we present some preliminary results about MADStat (paper counts, network centrality, journal ranking).  
In Section \ref{sec:application}, we analyze the text data in MADStat using Topic-SCORE as the main tool.  
In Section \ref{sec:extension}, we propose TR-SCORE (an extension of Topic-SCORE) for ranking different topics, and we also construct a cross-topic knowledge graph. Section \ref{sec:conclusion} contains a brief discussion.


\section{Topic Models and their Applications} 
\label{sec:topic} 
Topic model is one of the most popular models in text analysis.  
\cite{deerwester1990indexing} proposed the latent semantic indexing (LSI) 
as an ad-hoc approach to word embedding. Later,  \cite{Hof1999} proposed a probabilistic model for LSI, which is nowadays known as the topic model.  Hofmann's topic model can be described as follows. Given $n$ documents written with a vocabulary of $p$ words, let $X\in\mathbb{R}^{p\times n}$ be the {\it  word-document-count matrix}  where $X(j,i)$ is the count of the $j$th vocabulary word in document $i$. 
Write $X=[x_1,x_2,\ldots,x_n]$ so $x_i\in\mathbb{R}^p$ is the vector of word counts for document $i$.  
Suppose document $i$ has $N_i$ words.  For a weight vector (all entries are non-negative with a unit sum)  $\Omega_i \in \mathbb{R}^p$,     we assume 
\beq \label{pLSI0}
x_i\sim \mathrm{Multinomial}(N_i, \Omega_i), \qquad 1 \leq i \leq n. 
\eeq 
Here,  $\Omega_i$ is both the probability mass function (PMF) for $x_i$ and the vector of {\it population word frequency}; in addition, we implicitly assume the words are drawn independently from the vocabulary with replacement.  
%
%
%
%
%
%
Next, while there are a large number of documents, we assume there are only 
$K$ ``topics" discussed by these documents,   and $K$ is a relatively small integer.   
Fix $1 \leq i \leq n$ and consider document $i$. For a weight 
vector $w_i \in \mathbb{R}^K$ and PMFs $A_1, \ldots, A_K \in \mathbb{R}^p$,    we assume: (a) $w_i(k)$ is document $i$'s `weight' on topic $k$, $1 \leq k \leq K$, and (b) 
given that the document is (purely) discussing topic $k$, the population word frequency vector is $A_k$.  
Combining (a)-(b) and (\ref{pLSI0}), it is reasonable to assume $\Omega_i = \sum_{k=1}^n w_i(k) A_k$. 
%
%
%
%
%
%
%
%
%
%
Write $\Omega = [\Omega_1, \Omega_2, \ldots, \Omega_n]$, $A = [A_1, \ldots, A_K]$,  and $W = [w_1, w_2, \ldots, w_n]$. It follows that 
\beq \label{pLSI}
\Omega =AW.  
\eeq
We call $A$ and $W$ the {\it topic matrix} and the {\it topic weight matrix}, respectively. 

From time to time, we may normalize $X$ to the {\it word-document-frequency} matrix $D = [d_1,   \ldots, d_n] \in\mathbb{R}^{p\times n}$, where $D(j,i)=X(i,j)/N_i$ ($N_i$: total number of words in document $i$ as above).  
The primary goal of topic modeling is to estimate $(A, W)$ using  $X$ or $D$.

\subsection{Anchor words and identifiability of the topic model} \label{subsec:Anchor}  
We call a word {\it an anchor word of a given topic} 
if its occurrence almost always indicates that the topic is being discussed.
Consider the Associated Press (AP) \citep{harman1993first} data set for example. A pre-processed version of the data set consists of 2246 news articles  discussing three topics  
 ``politics", ``finance", and ``crime"  \citep{KaW2017}. 
 In this example, 
we may think ``gunshot" and ``Nasdaq" as anchor words for ``crime",  
 and ``finance", respectively.  In Model (\ref{pLSI0})-(\ref{pLSI}), we can make the concept more rigorously: we call word $j$ an {\it anchor word of topic $k$} if $A_k(j)\neq 0$ and $A_\ell(j)=0$ for all $\ell\neq k$.   
 
The notion of anchor word is broadly useful.   
First, it can be used to resolve the identifiability issue of the topic model. 
Without any extra conditions, Model (\ref{pLSI0})-(\ref{pLSI}) is 
non-identifiable (i.e.,  given an $\Omega$,  we may have multiple pairs of $(A, W)$ satisfying $\Omega = A W$).   To make the model identifiable, we may assume $\mathrm{rank}(W) = K$ and impose 
the {\it anchor-word condition}  (which requires that each of the $K$ topics has at least one anchor word).   The anchor-word condition was first proposed by \cite{arora2012learning} for topic models, which in turn was adapted from the separability condition \citep{donoho2003does} for nonnegative matrix factorization (NMF).   
Second, anchor words are useful in methodological developments: 
many topic modeling methods critically depend on the assumption 
that each topic has one or a few anchor words; for instance, see Section \ref{subsec:topic-SCORE} and 
\ref{subsec:anchor-recovery-methods} for descreptions of Topic-SCORE and anchor-word searching methods.  
Last but not the least, a challenge in real applications is that both the number of topics $K$ and the meanings of each estimated topics are unknown; 
we can tackle this with the (estimated) anchor words.  
See Section \ref{sec:application} for our 
analysis of the MADStat data for example,  where we use the estimated anchor words  to decide $K$, interpret each estimated topic, and assign an appropriate label for each of them.

\subsection{Topic-SCORE: A spectral approach to estimating the topic matrix $A$} \label{subsec:topic-SCORE}
In Hofmann's topic model (\ref{pLSI0})-(\ref{pLSI}), we can view 
$D = AW + (D - W) = ``\mbox{signal}" + ``\mbox{noise}"$, 
where (typically)  $\mathrm{rank}(AW) = K \ll \min\{n, p\}$. To estimate $A$ in such a ``low-rank signal matrix plus 
noise" scenario, it is preferable to emply a Singular Value Decomposition (SVD) approach, as SVD is effective 
in both dimension reduction and noise reduction. 

Topic-SCORE \citep{KaW2017} is an SVD approach  to topic modeling, consisting of   
  two main ideas: SCORE normalization and utilizing a  
low-dimensional simplex structure in the spectral domain. 
In detail,  \cite{KaW2017} pointed out that a prominent feature of text data is the {\it severe heterogeneity in word frequency}: 
the chance of one word appears in the documents may be hundreds of times larger than that of 
  another. This heterogeneity poses great challenges for textbook SVD approaches, 
so the vanilla SVD must be combined with proper normalizations.  \cite{KaW2017} proposed 
a pre-SVD approach, where for a diagonal matrix $M$ they constructed, 
they mapped the data matrix $D$ to $M^{-1/2} D$.  Unfortunately, while the pre-SVD normalization may reduce the effects of severe heterogeneity to some extent,  a major part of them persists.  To overcome the challenge,  
\cite{KaW2017} proposed a post-SVD normalization as follows. 
Let $\hat{\xi}_k$ be the $k$-th left singular vector of $M^{-1/2} D$. 
They normalized $\hat{\xi}_2, \ldots, \hat{\xi}_K$ by dividing each of them by 
$\hat{\xi}_1$ entry by entry. This gives rises to a matrix $\hat{R} \in \mathbb{R}^{n, K-1}$, 
where $\hat{R}(i, k) = \hat{\xi}_{k+1}(i) / \hat{\xi}_1(i)$ (by Perron's theorem \cite{HornJohnson},  all entries of $\hat{\xi}_1$ are positive under a mild condition).   
\cite{KaW2017} argued that, by combining the pre-SVD normalization and post-SVD normalizations, 
one can satisfactorily alleviate the effects of severe word-frequency heterogeneity.  The post-SVD normalization 
was inspired by the SCORE normalization (proposed by \cite{Jin2015} for analyzing network data with severe degree heterogeneity), thus the name Topic-SCORE.

\cite{KaW2017} discovered a low-dimensional simplex ${\cal S}$ with $K$ vertices as follows. 
For $1 \leq i \leq p$, let $\hat{r}_i'$ be the $i$th row of $\hat{R}$,  and view each $\hat{r}_i$ 
as a point in $\mathbb{R}^{K-1}$. They pointed out: (a) when word $i$ is an anchor word, 
then (up to small noise; same in (b)) $\hat{r}_i$ falls on one of the vertices of ${\cal S}$; (b) when word $i$ is a non-anchor word, $\hat{r}_i$ is in the interior of ${\cal S}$.  

This simplex structure reveals a direct relationship between $\hat{R}$ and $A$ 
($A$ is the quantity of interest) and gives rise to the Topic-SCORE approach as follows.   
Let $\hat{v}_1, \ldots,\hat{v}_K$ be the estimates of the vertices 
of ${\cal S}$. We can write each $\hat{r}_i$ uniquely as a convex linear combination of 
$\hat{v}_1, \ldots, \hat{v}_K$, with a barycentric coordinate vector $\hat{\pi}_i \in \mathbb{R}^{K}$.  
Topic-SCORE estimates $A$ by $\hat{A} = M^{1/2} \diag(\hat{\xi}_1) [\hat{\pi}_1, \ldots, \hat{\pi}_p]'$ (subject to a column-wise renormalization), where $\mathrm{diag}(\hat{\xi}_1)$ is the diagonal matrix whose diagonal entries are from $\hat{\xi}_1$.
In a noiseless case where $D = A W$,  \cite{KaW2017} showed that $\hat{A} = A$, 
so the approach is valid.  An interesting problem here is how to use the rows of $\hat{R}$ to estimate the vertices of ${\cal S}$ (i.e., Vertex Hunting (VH)).  This problem was studied in hyperspectral unmixing and archetypal analysis, with many available algorithms. \cite{KaW2017} recommended the sketched vertex search (SVS) algorithm \citep{JKL2017} for its superior numerical performance.  See \cite{SCOREreview} for more discussion on this. 

The major computation cost of Topic-SCORE comes from the SVD step, which can be excuted relatively fast. 
For this reason, Topic-SCORE is fast and can easily handle large corpora. For example, it takes only a minute to process the MADStat corpus in Section~\ref{sec:application}. Topic-SCORE is also theoretically optimal in a wide parameter regime \citep{KaW2017}.

\subsection{The anchor-word-searching methods for estimating $A$} \label{subsec:anchor-recovery-methods}  
\cite{arora2012learning,arora2013practical} proposed an anchor-word-searching approach which estimates $A$ by finding anchor words from the {\it word-word co-occurrence}  matrix $Q=DD'$. This method first normalizes each row of $Q$ to have unit-$\ell^1$-norm, with the resulting matrix denoted by $\bar{Q}$. It then applies a successive projection algorithm to rows of $\bar{Q}$, to get a subset $S\subset\{1,2,\ldots,p\}$ containing exactly one estimated anchor word per topic.  
The method then estimates $A$ by either a direct reconstruction or minimizing some objective functions (e.g., KL-divergence). 
\cite{arora2012learning,arora2013practical} are among the first works that utilize the anchor-word condition for topic modeling and provide explicit error rates.
A challenge it faces is that the rows of $\bar{Q}$ are in a very high-dimensional space.  
Similar to Topic-SCORE, their anchor-word-searching also relies on a $K$-vertex simplex, except for a major difference: this simplex is in $\mathbb{R}^p$ while the simplex in 
Section \ref{subsec:topic-SCORE} is in $\mathbb{R}^{K-1}$ (e.g., in the aforementioned AP dataset, $K = 3$, but $p$ is a few thousands).  
This gives Topic-SCORE an important edge (in both theory and  computation) when it comes to vertex hunting (VH) and subsequent steps of estimating $A$. 
In particular, Topic-SCORE improves the error rate in \cite{arora2012learning,arora2013practical}.

\cite{bing2020fast} proposed a different anchor-word-searching approach. Recall that $W \in \mathbb{R}^{K\times n}$ is the topic weight matrix; see Model (\ref{pLSI0})-(\ref{pLSI}).  Letting $\zeta_k=\|W_k\|_2/\|W_k\|_1$, where $W_k$ is $k$th row of $W$,  they assumed $\frac{W_k'W_\ell}{\|W_k\|\|W_\ell\|}<\frac{\zeta_k}{\zeta_\ell}\wedge \frac{\zeta_\ell}{\zeta_k}$, for $1\leq k\neq \ell\leq K$. For the same $\bar{Q}$ as above, let $S_i$ be the set of indices $j$ such that $\bar{Q}(i,j)$ attains the maximum value of row $i$.   
\cite{bing2020fast} proposed an approach and showed that if (a) the above assumption holds,  and (b) the model is noiseless (i.e., $D = AW$), then the approach can fully recover the set of anchor words from the index sets $S_1,S_2,\ldots,S_n$. 
Extending the idea to the real case (where $D = A W + ``noise"$), they obtained an estimate for the set of anchor words, and then a procedure for estimating  $A$. 

\subsection{Other approaches for estimating $A$: EM algorithm and NMF approaches} 
The EM algorithm is a well-known approach to fitting latent variable models. 
It was noted (e.g. \cite{mei2001note}) that Model \eqref{pLSI0}-\eqref{pLSI} is equivalent to a latent variable model, 
so we can estimate $A$ using the EM algorithm. 
Such an approach is interesting but faces some challenges. First,  it does not explicitly use any anchor-word condition, so the model being considered is in fact non-identifiable (see Section \ref{subsec:Anchor}).  
Also, since $\min\{n, p\}$ is typically large, the convergence of the EM algorithm remains unclear; even when the EM algorithm converges, the local minimum it converges to is not necessarily the targeted $(A,W)$ 
(which is uniquely defined under a mild anchor-word condition; see Section \ref{subsec:Anchor}). 

Also, note that Model (\ref{pLSI0})-(\ref{pLSI}) implies $D = A W + ``noise"$, where $(D, A, W)$ 
are all (entry-wise) non-negative matrices; hence, the problem of estimating $(A, W)$ can be recast as a non-negative matrix factorization (NMF) problem. There are many NMF algorithms (e.g., see \cite{gillis2013fast}), which are proved to be successful in applications such as image processing \citep{lee1999learning}, recommender systems, and bioinformatics.  However, 
a direct use of them in topic modeling faces challenges.   
The noise in most NMF settings is additive and homoscedastic, but the noise 
matrix $D-\mathbb{E}[D]$ in the topic model is non-additive and severely 
heteroscedastic, as indicated by the multinomial distribution. In Model (\ref{pLSI0})-(\ref{pLSI}),  the variance of $D(j,i)$ is proportional to word $j$'s frequency in document $i$.  Because of severe word-frequency heterogeneity, the variances of $D(j, i)$ may have different magnitudes, hence, a direct application of NMF algorithms often yields non-optimal error rates. 

\subsection{Estimating the topic weight matrix $W$} \label{subsec:estimateW}
In Model (\ref{pLSI0})-(\ref{pLSI}), $D = A W + ``noise"$,  and  both $A$ and $W$ 
are unknown. While most existing works focused on estimating $A$, $W$ is also of interest (e.g., see Section \ref{sec:application}).  
To estimate $W$,  a natural approach is to first obtain an estimate $\hat{A}$ for $A$, and 
then estimate $W$ by fitting the model $D = \hat{A} W +``noise"$.  Recall that $W = [w_1, \ldots, w_n]$. 
\cite{KaW2017} proposed a weighted least square approach, where for each $1 \leq i \leq n$,  
 it estimates $w_i$ by $\hat{w}_i=\mathrm{argmin}_{w}\|\Theta(d_i-\hat{A}w)\|^2$, with $\Theta\in\mathbb{R}^{p\times p}$ being a diagonal weight matrix (as $w_i \in \mathbb{R}^K$ and $K$ is typically small, this is 
is a low-dimensional regression problem).   To handle severe word-frequency heterogeneity, \cite{KaW2017} suggested  $\Theta=M^{-\frac{1}{2}}$, with the same $M$ as in Section~\ref{subsec:topic-SCORE}.    
For our study on the MADStat data in Section~\ref{sec:application}, we find that taking $\Theta=I_p$ also works fine, if a ridge regularization is added.  Noting that the word count vector $x_i$ is distributed as $\mathrm{Multinomial}(N_i, Aw_i)$, we can also estimate $w_i$ by some classical approaches, such as MLE, where we replace $A$ by $\hat{A}$ in the likelihood.

The above raises a question: Since $D = A W  + ``noise"$, 
can we first estimate $W$ and then use $\widehat{W}$ to estimate $A$? 
There are two concerns. 
First,   in some settings, the optimal rate for estimating $A$ is faster than that of estimating $W$ (see Section \ref{subsec:Rate}).  
Therefore, if we first estimate $W$ and then use $\widehat{W}$ to estimate $A$, then we may achieve the optimal rate in estimating $W$ but likely not 
in estimating $A$.   If we first estimate $A$ and then use $\widehat{A}$ to estimate $W$, we have optimal rates in estimating both.  
Second, many approaches for estimating $A$ rely on the assumption that each topic has some  
anchor words (see Sections~\ref{subsec:topic-SCORE}-\ref{subsec:anchor-recovery-methods}). 
If we extend them to estimate $W$, we need to similarly 
assume that each topic has some pure documents (document $i$  is  pure  
if $w_i(k) = 1$ and $w_i(\ell) = 0$ for $\ell \neq k$).   
However, in many  applications,  it is more reasonable to assume the existence anchor words than the existence of pure documents (especially when documents are long).  Therefore,  though the roles of  $A$ and $W$ may appear symmetrical to one other, 
they are not symmetrical in reality.

\subsection{The optimal rates for estimating $(A, W)$} \label{subsec:Rate}
For simplicity, as in many theoretical works on topic modeling,  we assume $N_1 = \ldots   N_n = N$; i.e., documents have the same length. We may have either a long-document (LD) case where $N/p=O(1)$ or a short-document (SD) case where $N/p = o(1)$ ($p$: size of the vocabulary).

Consider the rate for estimating $A$.   For any estimate $\hat{A}$, we measure the loss by  the $\ell^1$-error:   ${\cal L}(\hat{A},A)=\sum_{k=1}^K \|\hat{A}_k-A_k\|_1$ (subject to a permutation in the $K$ columns of $\hat{A}$). 
The minimax rate is defined as  $R_n = \inf_{\hat{A}}\sup_{A}\mathbb{E}{\cal L}(\hat{A},A)$. 
In the LD case, when $K$ is finite, $R_n \asymp \sqrt{p/(Nn)}$ up to a multi-$\log(p)$ factor (e.g., $\sqrt{\log(p)}$)  \citep{KaW2017}; when $K$ grows with $(n,p)$,  $R_n \asymp K\sqrt{Kp/(Nn)}$, also up to a multi-$\log(p)$ factor \citep{bing2020fast}. In the SD case, the optimal rate is unclear.  
Some minimax upper bounds were derived \citep{arora2012learning,KaW2017}, but they do not yet match the minimax lower bound. The difficulty of the SD case is that the majority of words have a zero count in most documents, which poses challenges in theoretical analysis.

Consider the rate for estimating $W$. Similarly,  for any estimate $\hat{W}$, we measure the loss by ${\cal L}(\hat{W},W)=\frac{1}{n}\sum_{i=1}^n\|\hat{w}_i-w_i\|_1$ (up to a permutation in the $K$ rows in $\hat{W}$) and define the minimax rate as $R_n = \inf_{\hat{W}}\sup_{W}\mathbb{E}{\cal L}(\hat{W},W)$. 
\cite{wu2022sparse} showed that $R_n \asymp \sqrt{K/N}$. In an apparently parallel work, \cite{klopp2023assigning} considered the Frobenius loss $n^{-1/2}\|\widehat{W}-W\|_F$ and showed that the minimax rate is $K\sqrt{1/N}$. 
The minimax rates are flat in $n$: This is not surprising, because the number of free parameters in $W$ is proportional to $n$.

%
%
%
%
%
%
%

\subsection{Estimating the number of topics $K$} \label{subsec:estimateK} 
Almost all topic learning algorithms assume $K$ as known a priori, but $K$ is rarely known in real applications. How to estimate $K$ is therefore a fundamental problem.  

To estimate $K$ in such a ``low-rank matrix plus noise" situation, 
a standard approach is to use the scree plot: 
for a threshold $t$, we estimate $K$ as the number of singular values of $X$ 
that exceed $t$. \cite{KaW2017} showed that this estimator is consistent, 
under some regularity conditions.  This method does not need topic model fitting and is fast and easy-to-use, but how to select a data-driven $t$ is an open question.  
Alternatively, one may select $K$ using BIC or other information criteria:  for each candidate of $K$, we obtain $(\hat{A}, \hat{W})$ by applying a topic learning algorithm, and estimate $K$ by the candidate that minimizes BIC. 
Also, alternatively, one may use the cross validation (CV) approaches, by estimating a topic model for each candidate $K$ and each training-validation split.  A commonly-used validation loss is the {\it perplexity}. It measures the predictive power of a trained language model on the held-out test set.  To use perplexity, we usually assume $w_i$ are iid generated, so the approach is more appropriate for the Bayesian version of the topic model to be introduced in Section \ref{subsec:LDA};   we can also use a full Bayesian approach by imposing a prior on $K$ and selecting $\hat{K}$ to minimize the marginal likelihood \citep{taddy2012estimation}.  In both the BIC and CV approaches, we need to fit the topic model many times, so the computational cost is high.  

In simulation studies,  it has been noted that (a) none of these methods is uniformly better than others, and which method is the best  depends on the data set,  and (b)  the popular perplexity approach often over-estimates $K$.  For these reasons, in real applications, whenever some inside information is available, we hope to use them to help determine $K$. For example, in the study of MADStat (see Section \ref{sec:application}), we investigate the estimated anchor words by Topic-SCORE for different $K$, and use our knowledge of the statistical community to choose the $K$ with the most reasonable results.  
In some applications, what the best $K$ is depends on the perspectives of the users,  and even experts may differ in their opinions.  In such a case, we may want to consider several 
different $K$. Such a flexibility may be helpful.

\subsection{Global testing associated with topic models}  
The problem of global testing is closely related to the problem of estimating of $K$.  
The goal is to test $H_0: K=1$ versus $H_1:K>1$.  
Global testing is a fundamental problem:  if no method can 
reliably tell between $K = 1$ and $K > 1$, it is merely impossible to estimate $K$ or estimate 
the matrices  $(A, W)$ in Model (\ref{pLSI0})-(\ref{pLSI}).     

Recall that $x_i \sim \mathrm{Multinomial}(N_i, A w_i)$, $1 \leq i \leq n$,  in Model (\ref{pLSI0})-(\ref{pLSI}).  
\cite{cai2023testing} proposed a test statistic $\psi_n$ called DELVE.  
They showed that when $K = 1$, although the model has many unknown parameters, 
$\psi_n \goto N(0,1)$, and the limiting distribution does not depend on unknown parameters. 
This result is practically useful. 
For example, we can use it to compute an approximate 
$p$-value and use the $p$-value to measure the 
research diversity of  different authors in the MADStat dataset;  see Section 3.3 of \cite{ji2022co} 
for a similar use of global testing in the network setting \citep{jin2018network,JKL2019}. 

Denote by $\lambda_2$ the second largest (in magnitude) eigenvalue of $\Sigma_A=A'[\diag(A{\bf 1}_K)]^{-1}A$.   
Similar as in Section \ref{subsec:Rate}, we assume $N_i = N$ for $1 \leq i \leq N$.      
Consider the DELVE test that rejects $H_0$ if $|\psi_n| \geq t$, for a threshod $t > 0$.    
\citep{cai2023testing} showed that this test achieves a sharp phase transition as follows.   
If  $|\lambda_2| / \sqrt{p/(Nn)}\to \infty$, for an appropriate $t$, the sum of the Type I and Type II errors of the DELVE test converges to $0$ as $p \goto \infty$.   
If $|\lambda_2| / \sqrt{p / (Nn)} \goto 0$, for any test, the sum of the Type I and Type II errors converges to $1$.  
Compared with earlier works (e.g., \cite{KaW2017,bing2020fast}), such a result is more satisfying. In earlier works,  
we usually assume all eigenvalues of $\Sigma_A$ are at the order of $O(1)$. Here, 
we may have $\lambda_2 = o(1)$, especially when $p \ll Nn$. 

\subsection{The latent Dirichlet topic model and its estimation} \label{subsec:LDA}
The latent Dirichlet allocation (LDA) model by \cite{BNJ2003}  is one of the most popular topic models, and 
it can be viewed as a Bayesian version of the Hofmann's topic model. 
In the LDA model, we start with Model \eqref{pLSI0}-\eqref{pLSI} and further assume that the topic weight 
vectors $w_1,w_2,\ldots,w_n$ are i.i.d. drawn from a Dirichlet distribution with parameters  $\alpha = (\alpha_1, \ldots, \alpha_K)$, where $\alpha_k \geq 0$ and $\sum_{k=1}^K \alpha_k = 1$. 
The LDA model has parameters $(A, \alpha)$ and 
treats $w_i$'s as latent variables.  In such a setting, $(A, \alpha)$   
are estimated by a variational EM algorithm, and the posterior of $w_i$'s can be obtained using MCMC. This is essentially the approach proposed by \cite{BNJ2003}.  
Compared to Model (\ref{pLSI0})-(\ref{pLSI}), 
LDA does not assume any structure on the topic matrix $A$. 
Therefore, if our goal is to estimate $A$, 
all those methods in Sections~\ref{subsec:topic-SCORE}-\ref{subsec:anchor-recovery-methods} are still applicable. 
In particular,   compared to the variational EM approach of \cite{BNJ2003},  Topic-SCORE in Section \ref{subsec:topic-SCORE} is not only faster but also provides desired theoretical guarantees \citep{KaW2017}.  
On the other hand,  LDA puts a Dirichlet prior on the topic weights $w_i$. This allows us to learn the posterior distribution of $w$ and may provide additional insights. 
Recall that in Section \ref{subsec:estimateW}, we have proposed a regression approach to estimating $W$ (without any priors on $W$).  
The regression approach is still useful for the LDA model (e.g., we can use this method to estimate 
the parameter $\alpha$ in the LDA model, and plug the estimated value to the variational EM algorithm).

\subsection{The $m$-gram topic models}
Hofmann's topic model and the LDA are so-called  {\it bag-of-word}  or {\it uni-gram} models, as they only model the counts of single words, neglecting word orders and word context.  There are several ideas about extending these models to incorporate word orders and word context.  

One idea is to simply expand the vocabulary to include phrases. For example, we may include all possible $m$-grams in the vocabulary  (an $m$-gram is a sequence of $m$ words).   
Unfortunately, even for a small $m$,  the size of this vocabulary 
is too large, making  topic estimation practically 
infeasible. To address the issue, we may only include 
a subset of carefully selected $m$-grams. 
For example, we may exclude low-frequency phrases or apply a phrase retrieval algorithm \citep{fagan1988experiments}.  
Once the vocabulary is determined, we treat each item in the vocabulary 
as a ``word" and model them  by (\ref{pLSI0})-(\ref{pLSI}) same as before; 
the resulting model is still a uni-gram model in flavor.

Another idea is the bigram topic model \citep{wallach2006topic}.  For each $1 \leq i \leq n$, document $i$ is modeled as an ordered sequence of words 
satisfying a Markov chain with a transition matrix $M_i \in \mathbb{R}^{p\times p}$ ($p$: vocabulary size),  where $M_i(j,\ell)$ is the probability of drawing word $\ell$ when the word immediately preceding it is  
word $j$.  For transition matrices $A_1, A_2, \ldots, A_K \in \mathbb{R}^{p\times p}$, 
$M_i=\sum_{k=1}^K w_i(k)A_k$, where each $A_k$ is treated as a ``topic" and $w_i\in\mathbb{R}^K$ is the topic weight vector as before. 
\cite{wallach2006topic} proposed a Gibbs EM algorithm for estimating the parameters and showed that, compared to the unigram topic model, this bigram model led to a better predictive performance and more meaningful topics on two real-world datasets. 


\subsection{Supervised topic models}
In many applications, we observe not only text documents but also some response variables associated with documents. For example, many online customer reviews contain numeric ratings;
we treat a review as a text document and the corresponding rating as the response. 
We would like to build a joint model for text and response, to help predict future ratings.  

The model in \cite{ke2019predicting} is a supervised topic model of this kind. 
This paper studied the problem of  how to use news articles to improve 
financial models. They focused on the news articles in Dow Jones Newswire. These articles are tagged with the identifier of a firm  
(the study excluded articles tagged with multiple firms). 
They model the news article with Model (\ref{pLSI0})-(\ref{pLSI}) and $K = 2$ (so there are only two topics),  
where the two topics are ``positive sentiment" and ``negative sentiment", respectively.    
In such a simple case, 
for any $1 \leq i \leq n$,  let $w_i = (a_i, 1-a_i)'$ be the topic weight of document $i$ as before ($w_i$ captures the ``sentiment" level of article $i$). Meanwhile, 
let $y_i$ be the stock return of the firm being tagged with document $i$.  
They assume that $\mathbb{P}(y_i > 0) = f(a_i)$ for an (unknown) function $f$ that is monotone increasing. 
This model jointly models text and return data, allowing for a better estimation of $w_i$ (which in turn may 
lead to a better prediction of stock returns). 
Compared with other approaches that also estimate news sentiment and use it
to predict returns, this approach 
has a substantial improvement on real-data performance.   
Moreover, see \cite{mcauliffe2007supervised} for other supervised topic models with a similar flavor.

\section{Deep neural network approaches to natural language processing} 
\label{sec:DNN}
The deep neural network approaches to natural language processing
(DNN-NLP)   
have become very popular recently, with successes observed 
in a variety of NLP tasks such as  text classification, question answering,   machine translation, among others \citep{otter2020survey}.  

In statistics, a ``model"  is a generative model with 
some unknown parameters we need to estimate. In DNN-NLP, 
researchers use the term ``model" slightly differently: 
a neural language model usually refers to a {\it pre-trained neural network} equipped with estimated parameters. 
A neural language model usually consists of three components as follows. 

\begin{itemize}
\item {\it A neural network architecture}. This is the core of a neural language model. It specifies how an input text is processed to generate the desirable output. The encoder-decoder structure is commonly used: the encoder is a neural network that maps the input text into a numeric vector (a.k.a., the encoder state), and the decoder converts the encoder state to the targeted output (e.g., a variable-length sequence of tokens).  
Many neural network models were inspired by new architectures proposed in the literature.  

\item {\it The NLP tasks used to train the neural networks}. A neural language model 
usually targets on one specific task (e.g., machine translation) or several specific NLP tasks (e.g., the BERT model \citep{devlin2018bert} outputs document embeddings, which can be used in various downstream tasks).  
In either case, pre-training the neural networks (i.e., estimating the parameters) must use specific NLP tasks to define the objective function. Hence, the same architecture may lead to different neural language models if they are pre-trained using different NLP tasks.  

\item {\it The text corpora and domain knowledge used in training}.  
Even with the same architecture and the same NLP tasks in training, the resulting neural language model still  varies with the training corpora. One strategy is selecting training corpora to obtain a domain-specific language model. For example, BERT has variants such as BioBERT \citep{lee2020biobert} trained using publications in biomedicine. Besides domain-specific corpora, other knowledge such as a domain-specific vocabulary can also be employed.  
\end{itemize}

The research on DNN-NLP has multiple goals,  including but not limited to  (a) Prediction of the next word given the previous words in a sentence (e.g., GPT family \citep{radford2018improving}),  (b) Extraction of numeric features from text (e.g.,  BERT family \citep{devlin2018bert}),  and (c) modeling the (synatic and semantic) relationships of words (e.g., word2vec \citep{mikolov2013efficient}).  
 DNN-NLP is a fast-developing area, which is hard to review comprehensively  (especially as our focus is on the topic modeling approaches and the 
MADStat data).    For these reasons, we select a few interesting topics in DNN-NLP to review,  focusing on 
(a) popular DNN architectures for NLP, (b) BERT, a powerful feature extraction tool developed by Google Inc. We also discuss word embedding and how to apply a neural language model (e.g., BERT) to a text corpus in our own research (see Remarks 1-2).

\subsection{Commonly used neural network architectures} \label{subsec:architecture}
Some well-known network architectures for NLP include the convolutional neural networks (CNNs), recursive neural networks (RNNs), and transformers.  CNNs and RNNs are more traditional, 
and transformers have become very popular in recent years.

CNNs use structural layers (e.g., convolutional layers and pooling layers) to capture the spacial patterns in the input, 
and are extensively used in signal (speech, image, video) processing.  
 In processing a text document,  sometimes it is not important whether certain words appear, but rather whether or not they appear in particular localities. Hence, 
 CNNs are also useful for NLP tasks such as sentence modeling  \citep{kalchbrenner2014convolutional} and sentiment analysis \citep{dos2014deep}. 
 
RNNs are especially useful for sequence data with variable-lengths, making them suitable for text analysis. The long short-term memory (LSTM) network \citep{hochreiter1997long}
is the most popular variant of RNNs.  
 In the vanilla RNNs, information may be diluted with successive iterations, preventing the model to ``remember" important information from the distant past. LSTMs add neurons (called ``gates") to retain, forget, or expose specific information, so it can better capture the dependence between two far-apart words in the sequence.  The standard LSTMs are unidirectional (i.e.,  text is processed left-to-right).  It is preferred to process text bidirectionally, as a word may depend on the words behind it. The bidirectional LSTMs combine outputs from left-to-right layers and right-to-left layers. 

The transformers  \citep{vaswani2017attention} are
a type of architectures based on the {\it attention} mechanism \citep{bahdanau2014neural}. 
In a traditional encoder-decoder pair, the encoder maps the input sequence into a fixed-length vector, and the decoder has access to this vector only. The attention mechanism allows the encoder to pass {\it all the hidden states} (not just the final encoded vector) to the decoder, along with annotation vectors and attention weights to tell the decoder which part of information to ``pay attention to".  
The attention mechanism was shown to be much more effective than RNNs in processing long documents. 
\cite{vaswani2017attention} proposed a special architecture called {\it transformer}  that uses self-attention within each of the encoder and decoder and cross-attention between them. The transformer 
has become the most popular architecture in NLP.  For example, the encoder part of the transformer is the building block of models like BERT (see below), and the decoder part of the transformer is the building block of models like GPT \citep{radford2018improving} for text generation.

\subsection{BERT} \label{subsec:BERT}   
The bidirectional encoder representations from transformers (BERT) is a state-of-the-art language model developed by Google AI Language \citep{devlin2018bert}, which provides a numerical representation for each sentence. 
As mentioned before,  a neural language model consists of three components: architecture, pre-training tasks, and training corpora.  For architecture, BERT uses the transformer encoder with bi-directional self-attention. 
For training corpora, BERT uses the BooksCorpus (800M words) \citep{zhu2015aligning} and English Wikipedia (2,500M words).  The main innovation of BERT is in the pre-training tasks it used: BERT was pre-trained using two tasks, the masked language modeling (MLM) and next sentence prediction (NSP). 
In MLM, some tokens of the input sequence are randomly masked, and the objective is to predict those masked tokens from their left and right contexts. In NSP, the input are two sentences A and B from a corpus, and the objective is to tell if B is the next sentence of A. These tasks do not require manual labeling of text.  

BERT has been applied to different downstream NLP tasks, with superior performances. Numerous language models have been created based on BERT, such as modifications of the architecture (e.g., ALBERT and DistillBERT) and pre-training tasks (e.g., RoBERTa and ELECTRA), adaptation to other languages (e.g., XLM and ERNIE), and inclusion of domain-specific corpora (e.g., BioBERT and UmlsBERT). See \cite{rahali2023end} for a comprehensive survey. 


{\bf Remark 1}.  Another major goal of NLP is to learn the syntactic and semantic relationships 
between words. To do this, a standard approach is word embedding (i.e., find vector representations of words).  
Despite the fact that word embedding is frequently used in neural language models (often as the first layer),   
its primary purpose is to understand or mimic various syntactic and semantic regularities in natural 
languages. 
%
%
%
 A frequently mentioned example is that vector(``king") $-$ vector(``man") $+$ vector(``woman") $\approx$ vector(``queen").    Word2vec \citep{mikolov2013efficient} is a popular word embedding model.  
%
%
It was trained using a Google News corpus, and its performance was tested on a semantic-syntactic relationship question set manually created by the authors.

%
%
%
%

{\bf Remark 2}.  Many modern DNN-NLP tools (such as BERT)  are owned by 
high-tech companies. 
They were trained with a huge amount of data and efforts, and many parts of them are not publicly available.  A typical NLP user has his/her own (domain-specific) text corpus  (1K to 10K documents), which are not large enough to re-train BERT (say). To help these users to apply modern DNN-NLP tools,  there are two approaches:  transfer learning and fine tuning. 
In the first approach, 
the user inputs his/her own documents to the BERT (say) and obtain 
an embedded vector for each document. The 
embedded vectors can then be used as features for downstream analysis.  In the second approach, a user may alter the parameters of the pre-trained model. By adding additional layers to the neural networks, one can convert the output of a pre-trained neural language model to the targeted output of a downstream task (e.g., document classification). Next, all the parameters---those in the pre-trained model and those for the added layers---are updated together (this can done by running stochastic gradient descents starting from parameters of the pre-trained models).

\section{MADStat basics: paper counts, journal ranking, and network centrality} 
\label{sec:basic}   
The {\it multi-attribute dataset on statisticians} (MADStat) contains the bibtex (e.g., author, title, abstract, journal, year, references, etc.) and citation information of  83,331 papers from 47,311 authors, spanning $41$ years (1975-2015). We collected and cleaned the dataset with substantial time and efforts and have made it publicly available (the links to download the dataset can be found in \cite{ji2022co}). In the supplementary material, we present  
(a) details on data collection and cleaning, (b) the list of the $36$ journals and their abbreviations, and 
(c) supplementary results of the text analysis conducted in this paper (such as selection of $K$ for Topic-SCORE). 
In this section, we discuss some basic findings on the data set, including paper counts, 
network centrality, and journal ranking.

\subsection{Paper counts} 
\begin{figure}[htb!]  
		\centering   
                 \includegraphics[height=.26\textwidth]{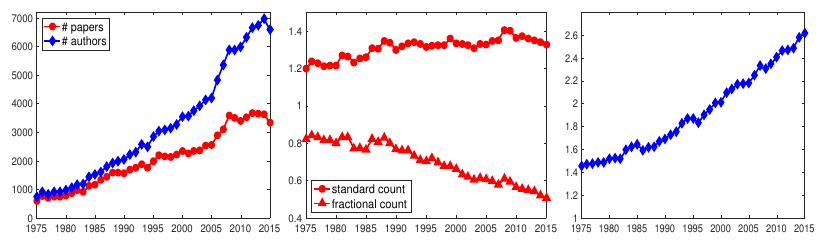}
		\caption{Left: total numbers of papers and active authors in each year. Middle: average number of papers per author in each year. Right: average number of authors per paper in each year.}\label{fig:productivity1} 
\end{figure}

The paper counts provide valuable information for studying how the productivity of 
statisticians evolve over time.   In the left panel of  Figure \ref{fig:productivity1}, the red curve presents the 
number of papers per year and the blue curve presents  
the number of  {\it active}  authors per year (an author is active in a given year if he/she publishes at least 1 paper in that year). In both curves, we notice a sharp increase 
near 2005-2006, possibly because several new journals ({\it AoAS, Bay,  EJS})  
were launched between 2006 and 2008; see Table \ref{tab:journal} of the supplementary material. 
The middle panel of Figure \ref{fig:productivity1} presents the yearly paper counts, 
defined as the average number of papers per active author. We consider both standard count and fractional count, 
where for an $m$-author paper, each author is counted as published $1$ and $1/m$ papers, respectively.  In the standard count, the yearly paper counts increase between 1975 and 2009, from about $1.2$ paper per author to about $1.4$ paper per author, and decrease  after 2009, to about $1.3$ paper per author in 2015. In the fractional count, the yearly paper counts always decrease,  from about  $0.85$ paper per author in 1975 to about $0.5$ paper per author in 2015.  This can be explained by that 
the average number of authors per paper has been steadily increaseing over the years. 
See the right panel of Figure~\ref{fig:productivity1}, where we present the average number of authors per paper; the curve is seen to be steadily increasing.

The above counts can be further explained by 
Figure~\ref{fig:productivity2} of the supplementary material, in which (a) the paper count each year is partitioned into the counts of $m$-author papers for different $m$ and (b) the author count each year is partitioned into the counts of $k$-year-senior author for different $k$. The results show some interesting patterns, and we refer the readers to Section~\ref{sec:paper-counts} of the supplementary material for details.

%

\subsection{Network centrality} \label{sec:network}  
Network centrality (e.g., most-collaborative authors) 
provides information for the leadership and trends in statistical research. 
Table \ref{tab:topauthors} presents the top 10 authors who have the most coauthors, the most citers (a citer for any given author is any other author who has cited this author), and the most citations, respectively.  Table~\ref{tab:toppapers} (Appendix, Section \ref{sec:center}) presents the top 10 most-cited papers. Note that the numbers of coauthors, citers, and citations here are all counted using {\it only} the papers in our data range, so there may be some biases in our ranking. For example, in Table~\ref{tab:toppapers}, if we instead use the citation counts by Google Scholar on December 31, 2022, then the papers {\it  Benjamini $\&$ Hochberg (1995)} on FDR,  {\it  Donoho $\&$ Johnstone (1994)} on wavelets, and {\it Efron et al. (2004)} on LARS will receive better rankings,   as these papers have  many citations from papers outside our data range. 
Despite this, our approach is still valuable. For example, using our data, we can provide the ranking (e.g., by number of citations) for any author or any paper in our data set, but how to do this using Google Scholar is unclear: We need to build a large database for the citation relationships between many authors and papers and spend substantial time cleaning such citation data. Compared to Google Scholar, our citation data are of higher quality, so our results on network centrality shed new light that Google Scholar cannot provide.


\begin{table}[htb!]
\centering
	\caption{The top $10$ authors ordered by the number of coauthors, citers, and citations, respectively (we only count co-authors and citations within the range of MADStat).}
	\label{tab:topauthors} 
	\scalebox{0.8}{
		\begin{tabular}{lc|lc|lc}  
			Author name                      &  $\#$Coauthors   & Author name               & $\#$Citers  &  Author name              &  $\#$Citations  \\ 
			\hline
			Raymond Carroll            & 234  & Donald B. Rubin           & 5337  & Peter Hall        & 6847     \\
			Peter Hall                 & 222  & Nan Laird     & 5079  & Donald B. Rubin        & 6825     \\
			N. Balakrishnan & 186  & Bradley Efron       & 4500  & Jianqing Fan & 5726     \\
			Jeremy Taylor              & 159  & Robert Tibshirani   & 4076  & Robert Tibshirani         & 5074     \\
			Joseph Ibrahim             & 158  & Peter Hall          & 3789  & Nan Laird &     5040 \\
			Geert Molenberghs          & 146  & Arthur P. Dempster  & 3406  & Bradley Efron     & 4589     \\
			James S. Marron            & 130  & Scott Zeger         & 3311  & Raymond Carroll   & 4415     \\
			Malay Ghosh          & 119  & Kung Yee Liang       & 3231  & Scott Zeger       & 3802     \\
			Emmanuel Lesaffre                & 119  & Trevor Hastie     & 3174  & Trevor Hastie     & 3582     \\
			Xiaohua Zhou           & 119  & Raymond Carroll      & 3110  & Kung Yee Liang     & 3366     \\
		\end{tabular}
	}  
\end{table}

\subsection{Citation patterns and the sleeping beauties}

\begin{figure}[htb!]
	\centering
         \includegraphics[width=1\textwidth]{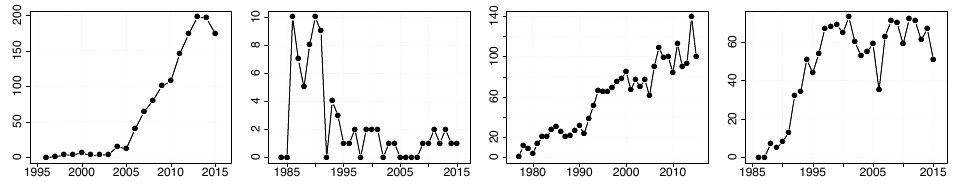}
	\caption{\small Yearly citation curves  for $4$  papers. Left to right:  ``sleeping
		beauty" ({\it Tibshirani (1996)} on Lasso), ``transient", ``steadily increasing" ({\it Dempster, Laird and Rubin (1977)} on EM algorithm),   and ``sudden fame" ({\it Liang and Zeger (1986)} on GLM).}
	\label{fig:4_pattern}
\end{figure}

Identification of representative citation patterns is an interesting problem, as it  helps distinguish short-term citation effects from long-lasting citation effects.
By a careful study of the yearly citation curves of individual papers, we
identify four representative citation patterns: ``sleeping beauty,'' ``transient,'' ``steadily increasing,'' and ``sudden
fame.'' ``Sleeping beauty" refers to the papers that receive low citations within a few years after publication but become frequently cited after a certain point (a.k.a. ``waking up''). Representative papers include the lasso paper, {\it Tibshirani (1996)}, and the FDR paper, {\it Benjamini and Hochberg (1995)}. 
``Transient'' refers to the papers that receive a good
number of citations for a few years shortly after publication, but then their citations drop sharply and remain low for years. ``Steadily increasing'' refers to those papers whose citations have been increasing at a modest rate for many years, with a large number of citations over a relatively long time period. 
Representative papers include {\it Dempster et al. (1977)} on   EM algorithm. 
``Sudden fame'' refers to papers
that receive a large number of citations shortly after publication and the citations remain high for many years. 
Representative papers include {\it Liang and Zeger (1986)} on longitudinal data,  {\it Gelfand and Smith (1990)} on  marginal densities,  and {\it Efron et al. (2004)} on LARS.  
See Figure \ref{fig:4_pattern}. 

The ``sleeping beauty" pattern is especially interesting. To identify the sleeping beautifies in our data range, we use the metric suggested by \cite{ke2015defining}. It outputs a measure $B_i$ for each paper $i$ (the details are in the supplementary material); the larger $B_i$, the more likely this paper is a sleeping beauty. 
We select the 300 papers with the largest maximum number of yearly citations and arrange them in the descending order of $B_i$. 
Table~\ref{tab:sb} and Figure \ref{fig:sb} in the supplementary material show the papers with largest $B_i$, such as {\it Tibshirani (1996)}, {\it Azzalini (1985)}, {\it Hubert $\&$ Arabie (1985)}, {\it Hill (1975)}, {\it Marcus et al. (1976)}, {\it Lunn et al. (2000)}, {\it Rosenbaum $\&$ Rubin (1983)}, {\it Bai \& Saranadasa (1996)}, {\it Holm (1979)}, {\it Clayton (1978)}, and {\it Fan $\&$ Li (2001)}.



\subsection{Journal ranking} \label{subsec:topic-JR}  
\begin{figure}[tb!]
	\centering
	\includegraphics[width=0.6\textwidth, height = 0.38\textwidth]{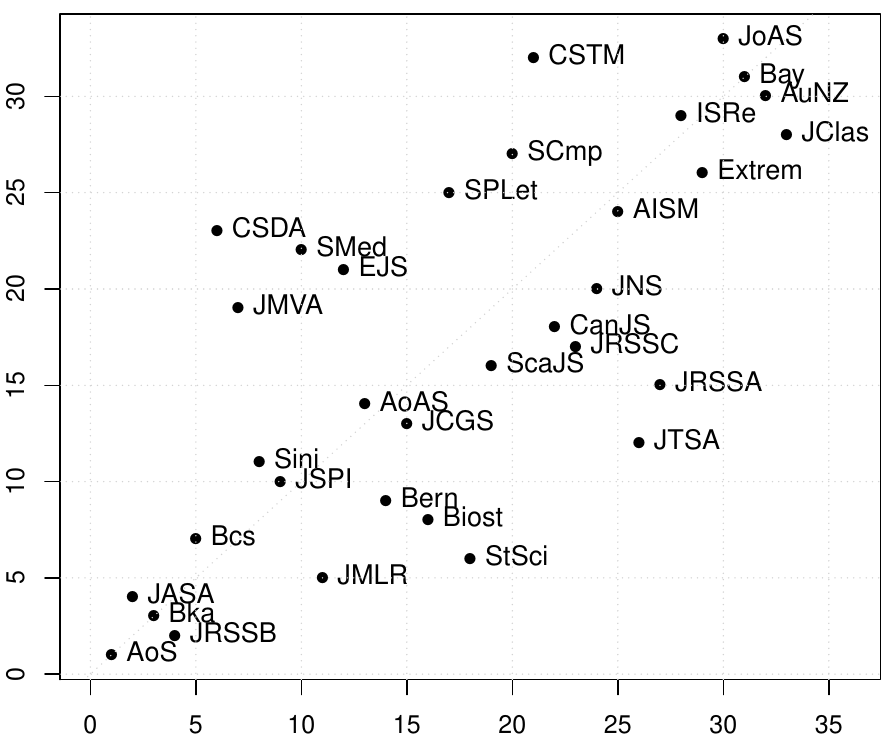}
	\caption{\small Journal ranking. 
	Each point is a journal (x-axis: ranking by PageRank, y-axis: ranking by Stigler's model). See Table~\ref{tab:journal} of the supplement for the full journal names.}
	\label{fig:rank1}
\end{figure}

Journal ranking has been widely used in appointing to academic positions, awarding research grants and ranking universities and departments.  
A common approach is the Impact Factor (IF),  
but IF is known to have some issues \citep{Va2016}. 
We instead use the Stigler's model \citep{St1994} for journal ranking:
Given $N$ journals, let $\mu_1,\ldots, \mu_N\in\mathbb{R}$ be their export scores; 
for two papers $i$ and $j$ published in journal $\ell$ and $m$, respectively, let $C_{ij}$ be the indicator of a citation from $i$ to $j$. We assume $\mathbb{P}(C_{ij} = 1 | C_{ij} + C_{ji} = 1) = 
\mathrm{exp}(\mu_{\ell} - \mu_m) / [1 + \mathrm{exp}(\mu_{\ell} - \mu_m))]$. 
We fit this model using the quasi-likelihood approach in \cite{Va2016}. For comparison, we also consider the PageRank approach (with the same tuning parameter $\alpha$ as suggested in \cite{Va2016}). 
Among the 36 journals (see Table~\ref{tab:journal}), there are relatively few citation exchanges between the 3 journals focusing on probability and the other 33 journals, so we exclude these 3 probability journals. For each journal pair, we count the citations between them using a 10-year window.  For instance, if 2014 is the ``current year," then we count one citation from journal $i$ to journal $j$ if and only if a paper published in journal $i$ in 2014 has cited a paper published in journal $j$ between 2005 and 2014. This gives rise to a $33 \times 33$ between-journal citation matrix for 2014.  Last, we take the sum of the two matrices for 2014 and 2015 to improve the stability and reliability of results. This is the final data matrix fed into journal ranking. 
The results are in Figure \ref{fig:rank1}. 

Both approaches rank {\it AoS}, {\it Biometrika}, {\it JASA}, and {\it JRSSB} as the top four. In particular, both approaches rank {\it AoS} as number 1 and {\it Biometrika} as number 3; PageRank ranks {\it JASA} as number 2 , and the Stigler approach ranks {\it JRSSB}  as number 2. The rankings of two methods are quite consistent with each other. A few exceptions are {\it CSDA}, {\it EJS}, {\it JMVA}, {\it JRSSA}, {\it JTSA}, and {\it SMed}. 
We notice that PageRank weighs each citation equally, while the Stigler model gives citations from higher-ranked journals larger weight than those from lower-ranked journals \citep{Va2016}. 
The results of PageRank are fairly close to that of ranking by citation numbers, but the results of the Stigler approach may be significantly different. A closer look at the citation counts reveals that a large proportion of citations of {\it SMed}, {\it CSDA}, {\it JMVA}, and {\it EJS} are self-citations, and after these self-citations are excluded, most citations to these journals are from journals with relatively low rankings. This explains why these journals are ranked relatively high by PageRank but relatively low by Stigler's model.  Also,  while neither {\it JTSA} nor {\it JRSSA} has a large number of citations, most of their citations come from journals with high rankings; consequently, the two journals are ranked much higher by Stigler's model than by PageRank. 

\section{Application of Topic-SCORE to the MADStat data set}  \label{sec:application}
In this section, we apply Topic-SCORE (see Section~\ref{subsec:topic-SCORE}) to analyze the abstracts in MADStat.  
We use all paper abstracts for the time period of 1990-2015 in $33$ journals, excluding the 3 probability journals {\it AIHPP},  {\it AoP}, and {\it PTRF} (see Table~\ref{tab:journal} about the full journal list), since the topics in these   journals are very different from in the other $33$ journals. This gives a total of $63,187$ abstracts.  We then perform a word screening by removing stop words and infrequent words, which gives rise to a vocabulary of $2,106$ words. 
Finally, we compute the length of each abstract by the number of words (a word not in the aforementioned vocabulary is not counted) and remove approximately the $10\%$ shortest abstracts. We have $56,500$ remaining abstracts. The details of pre-precessing are in Section~\ref{sec:text-preprocessing} of the supplementary material. The final data matrix is $X =[x_1,\ldots, x_n]\in\mathbb{R}^{p\times n}$ with $(p,n) = (2106, 56500)$; same as in Section~\ref{sec:topic}, $x_i\in\mathbb{R}^p$ contains the word counts of the $i$th paper abstract.    

\subsection{Anchor words and the $11$ identified topics} 
To apply Topic-SCORE, we need to decide the number of topics. This is a hard problem (see Section~\ref{subsec:estimateK}) and 
we tackle it by combining the scree plot, substantial manual efforts, and our knowledge of the statistical community (see Section~\ref{sec:SelectK} of the supplementary material). We find that that $K=11$ is 
the most reasonable choice.

\begin{figure}[htb!]
	\centering
	\includegraphics[width = 1\textwidth]{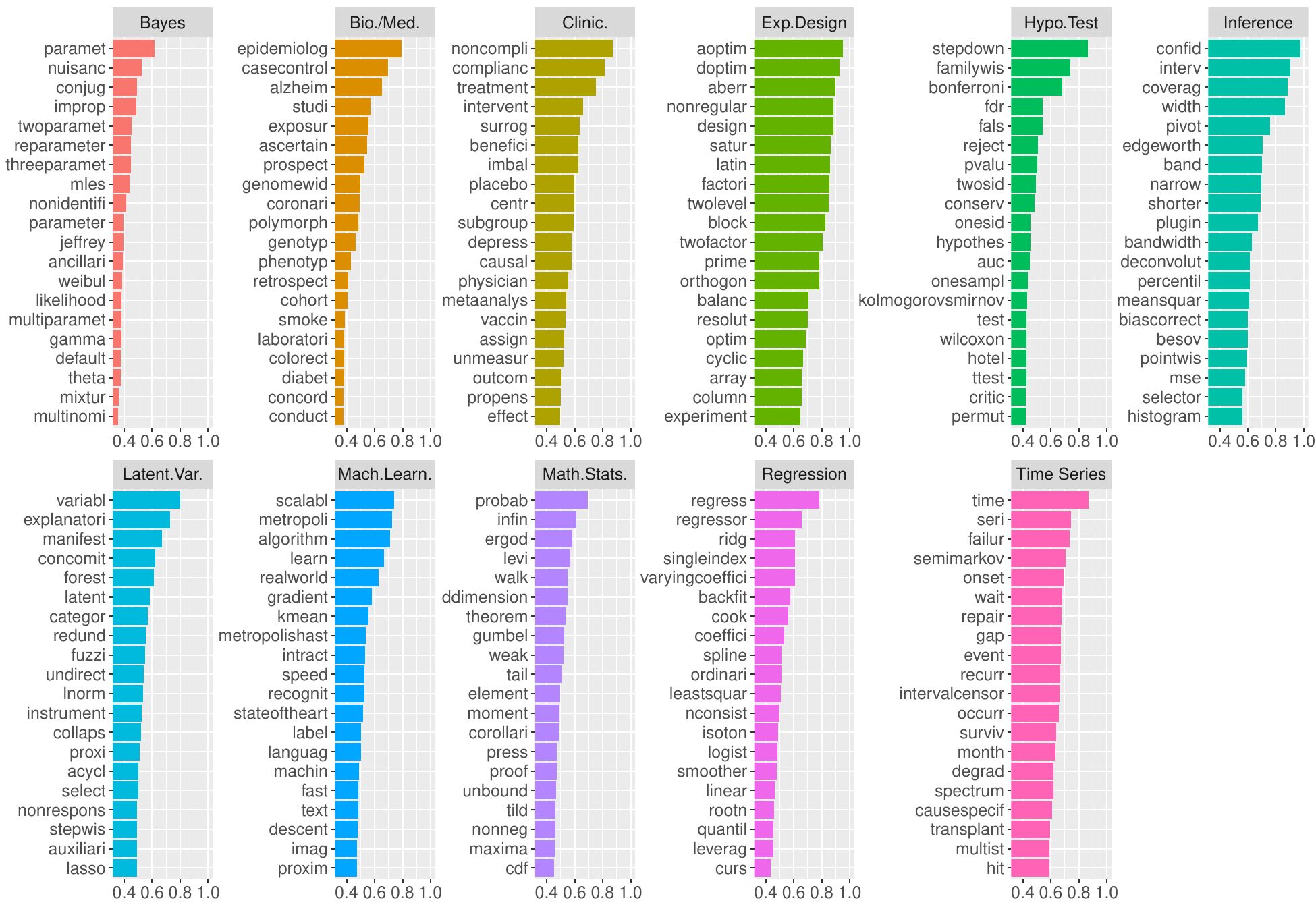}
	\caption{\small For $1 \leq k \leq K$ (where $K = 11$), Panel $k$ is the barplot of the 20 words $j$ that have the largest weight $a_j(k)$ among all words (the length of each bar is the value of $a_j(k)$).}
	\label{fig:anchor_words}
\end{figure}

Since $K = 11$, there are $11$ discovered research topics by Topic-SCORE. To interpret and label these topics, we introduce a rule for selecting `representative' words and papers for each topic. The anchor words (see Section~\ref{subsec:Anchor}) 
appear only in one topic. For example, ``lasso'' and ``prior'' may be anchor words for the topics  of ``variable selection'' and ``Bayes'', respectively.  Given    
$\hat{A}$, define the {\it topic loading vector} $a_j \in \mathbb{R}^{K}$ for each word $j$ by $
a_j(k) = \hat{A}_k(j)/[\sum_{\ell=1}^K \hat{A}_\ell (j)]$,  $1 \leq k \leq K$.    
Note that $0 \leq a_j(k) \leq 1$ and in theory $a_j(k) = 1$ if and only if word $j$ is an anchor word of topic $k$. 
Fix $1 \leq k \leq K$.  The most frequent anchor word in topic $k$ is the word $\hat{j}$ where 
$\hat{j} = \mathrm{argmax}_j \{a_j(k): 1 \leq j \leq p\}$. Similarly, we can define the $m$-th most frequent anchor word for any $m \geq 1$.  Figure \ref{fig:anchor_words} shows the $20$ most frequent anchor words for each of the 11 estimated topics. Based on these words, we suggest a name for each topic as in the second column of 
in Table \ref{tab:11-topic}.
To check if the proposed labels are reasonable and get more insight 
of each topic, we also use $\hat{W}$ to identify representative papers. 
For each $1 \leq k \leq 11$,  we pull out the top 300 papers with the largest $\hat{w}_i(k)$ (the titles of top-3 within each topic is given in Table~\ref{tab:repre_papers} of the supplementary material).  
We manually review the titles of these papers and come up with a list of suggested research topics umbrellaed by each of the brief topic label.  See the third column of Table \ref{tab:11-topic}.

\hspace*{-3cm}\begin{table}[tb!]
	\centering 
		\caption{Interpretation of the 11 estimated topics.}
	\scalebox{0.68}{
		\begin{tabular}{c| P{1.33 in}|p{.7in}| P{4.8in}}  
		\toprule
			& Topic Label & Abbreviation & Corresponding Research Topics\\
			\hline
			1& Bayesian statistics &  Bayes  & Bayesian methods \\
			2& Bio \& medical statistics & Bio/Med.  & Observational studies, genetics, genomics\\
			3& Clinical trials & Clinic. & Clinical trials, causal inference\\
			4& Experimental design & Exp.Design & Experimental design \\
			5& Hypothesis testing & Hypo.Test  & Hypothesis testing, goodness of fit\\
			6& Statistical inference &Inference& Confidence intervals, bootstrapping, empirical likelihood\\
			7& Latent variables &Latent.Var. & Latent variable model, incomplete data, mixtures, clustering, factor model, graphical model, variable selection, categorial data analysis, dimension reduction  \\ 
			8 & Machine learning & Mach.Learn. & Machine learning, computation, EM algorithm, Monte Carlo methods, clustering \\
			9& Mathematical statistics & Math.Stats. & Asymptotics, mathematical statistics, probability, stochastic process \\
			10& Regression analysis & Regression  & Linear models, nonparametric regression, quantile regression, semi-parametric models \\
			11& Time series & Time Series  & Time series, longitudinal data, stochastic processes, survival analysis\\
			\bottomrule
		\end{tabular} 
	}  
	\label{tab:11-topic} 
\end{table} 

Our topic learning results are  based on abstract similarity (i.e., the research areas covered by the same topic have similar word counts in their abstracts). Such a similarity does not necessarily imply the similarity in the intellectual content of the paper.  Also, our goal here is to use statistical   methods to identify a few interpretable topics, and it is possible that some research topics in the data set are not well represented here.

\subsection{Topic weights for representative authors}      
How to estimate the {\it research interests} of an author is an interesting problem. It helps us understand an author's research profile and may be useful in decision making (e.g., award,  funding, promotion); it may also help this author to plan for future research. 
We estimate the research interest of an author as follows. 
For an author $a$, let ${\cal N}_a\subset\{1,2,\ldots,n\}$ be the collection of papers he/she published   
in our data range. Each paper $i$ has an estimated 
topic weight vector $\hat{w}_i$ for its abstract.  A reasonable metric of author $a$'s interest on topic $k$ is $\bar{w}_a(k)=\frac{1}{|{\cal N}_a|}\sum_{i\in {\cal N}_a}\hat{w}_i(k)$, $1\leq k\leq 11$. Let $\bar{w}(k)$ be the average of $\hat{w}_i(k)$ over all 56,500 abstracts. We define the {\it centered topic interest vector} of author $a$ by $
z_a = \bar{w}_a - \bar{w}\in\mathbb{R}^{11}$.  The entries of $z_a$ sum to $0$, so it has both positive and negative entries. We are interested in its positive entries, since $z_a(k) > 0$ indicates a greater-than-average weight on topic $k$.

\begin{figure}[tb!]
	\centering
	\includegraphics[width=0.88 \textwidth]{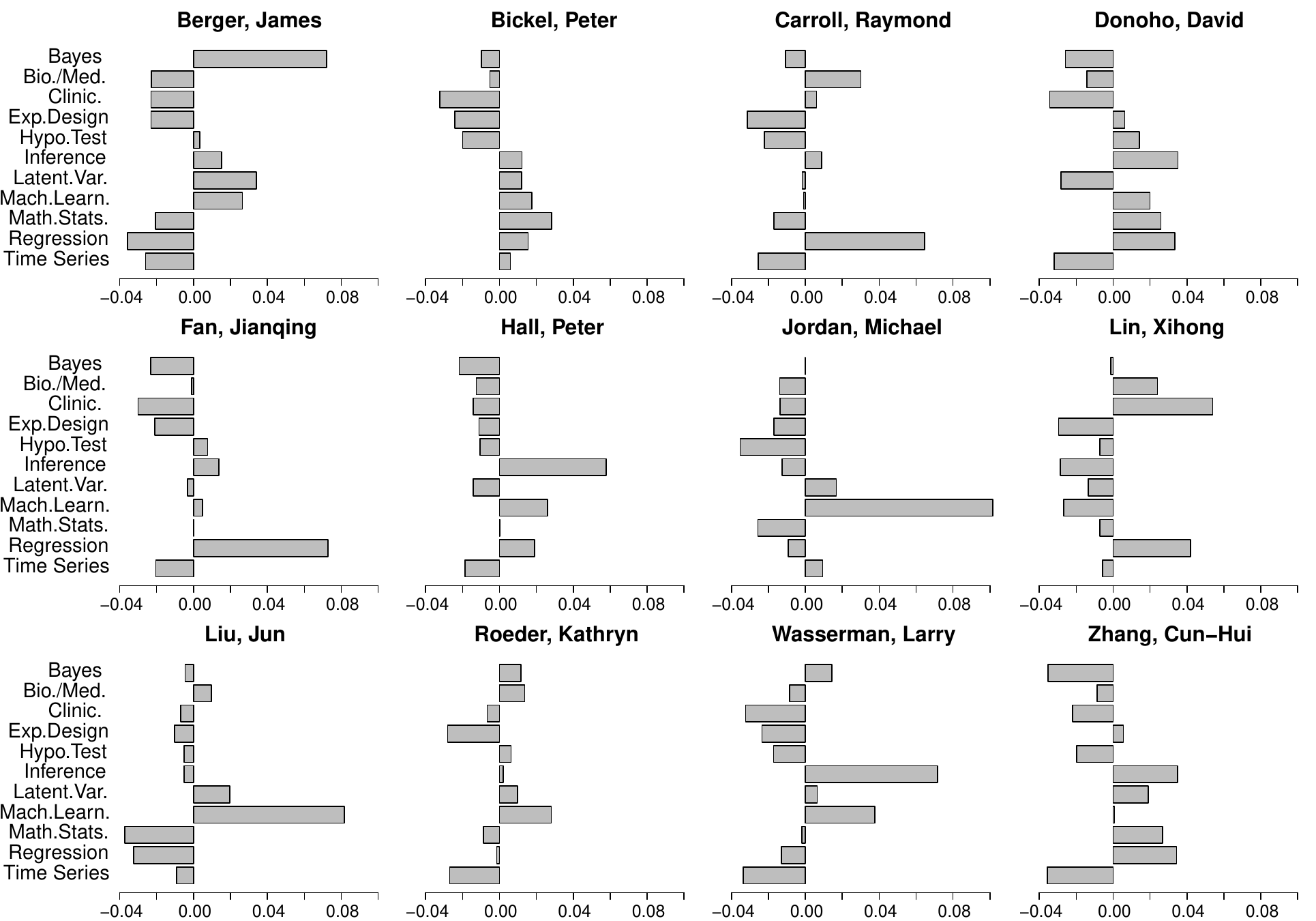}
	\caption{\small The overall topic interests of some authors. For interpretation purpose, we select 
		some authors we are familiar with, but similar figures can be generated for other authors.}
	\label{fig:tw_author_all}
\end{figure}

We can compute the vector $z_a$ for almost every author in our data range. 
Table~\ref{tab:tw_80authors} of the supplementary contains the results of 80 selected authors. 
Figure~\ref{fig:tw_author_all} presents $z_a$ for 12 representative authors. We have some interesting findings. 1) James Berger has a prominently high weight on {\it Bayes}; Raymond Carroll and Jianqing Fan have prominently high weights on {\it Regression}; and Michael Jordan and Jun Liu have prominently high weights on   {\it Mach.Learn.}  These results are reasonable: Berger has many works in Bayesian statistics and decision theory; Carroll has many works in semiparametric models; 
Fan has many works in nonparametric regression and high dimensional variable selection; Jordan  has many works in machine learning, nonparametric Bayes,  and Bayesian computation; and Liu has many works in Bayesian computation and MCMC. 2) 
Peter Hall has notably high weights on {\it Inference}, {\it Mach.Learn.}, and {\it Regression}; Xihong Lin has notably high weights on {\it Clinic.}, {\it Regression}, and {\it Bio./Med.}; Larry Wassermann has notably high weights on {\it Inference}, {\it Mach.Learn.}, and {\it Bayes}; and Cun-Hui Zhang has notably high weights on {\it Inference}, {\it Regression}, and {\it Math.Stat.}.  3) Figure~\ref{fig:tw_author_all} suggests that the research interests of Peter Bickel, David Donoho, and Kathryn Roeder are relatively diverse, covering many topics; these are consistent with  our impression of these authors and the information of 11 topics in Table~\ref{tab:11-topic}.

\subsection{Topic trends} \label{subsec:topic-TT} 
How to characterize the evolvements of  statistical research 
over time is an interesting problem \citep{KolarTaddy}. 
We tackle it by combining the estimated topic weights and the time and journal information of each paper.  

\begin{figure}[htb!]
	\centering
	\includegraphics[width = .8\textwidth, trim=0 15 0 0, clip=true]{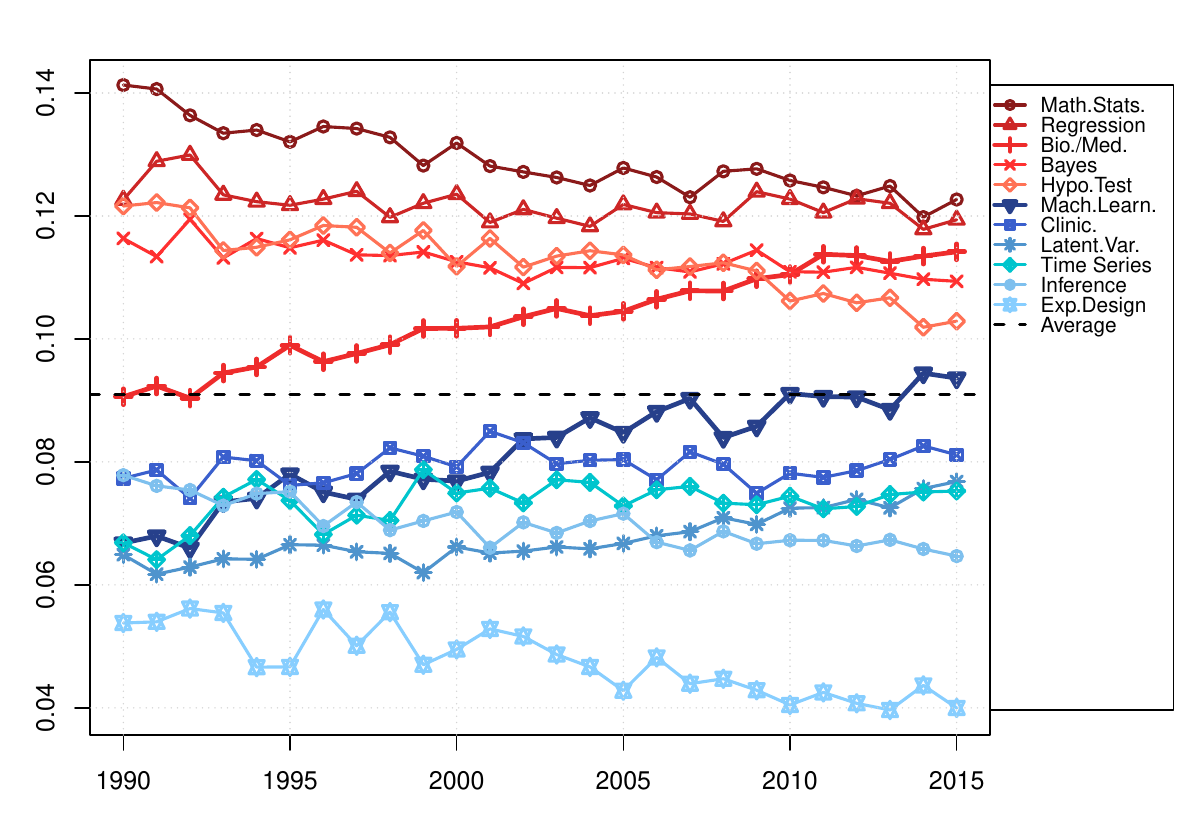}
	\caption{\small The yearly average topic weights (averaged for all 33 journals),  1990 -- 2015.} \label{fig:tw_dynamic_all}
\end{figure}  

First, we study how the yearly average topic weights change over time. 
Recall that $\hat{w}_i$ is the estimated topic weight vector for paper $i$ by Topic-SCORE. 
For each year,  we compute the average topic weight for {\it all papers published in this year},    
smoothed by a weighted moving average in a 3-year window (weights: $0.25$, $0.50$, and $0.25$).  
See Figure~\ref{fig:tw_dynamic_all}. We observe that 
the 5 topics, {\it Math.Stat.}, {\it Regression}, {\it Bio./Med.}, {\it Bayes}, and {\it Hypo.Test}, have higher-than-average weights, suggesting that they have attracted more attention;
from 1990 to 2015, the weight of {\it Bio./Med.} increases relatively fast, the weights of {\it Math.Stat.} and {\it Hypo.Test} gradually decrease, and the weights of {\it Regression} and {\it Bayes} are relatively flat. 
Among the remaining 6 topics, {\it Mach.Learn.} increases quickly; its weight has passed the overall average starting from 2014 ({\it Latent.Var.} is another topic where the weight is steadily increasing).

\begin{figure}[htb!]
\centering  
\includegraphics[width=1\textwidth]{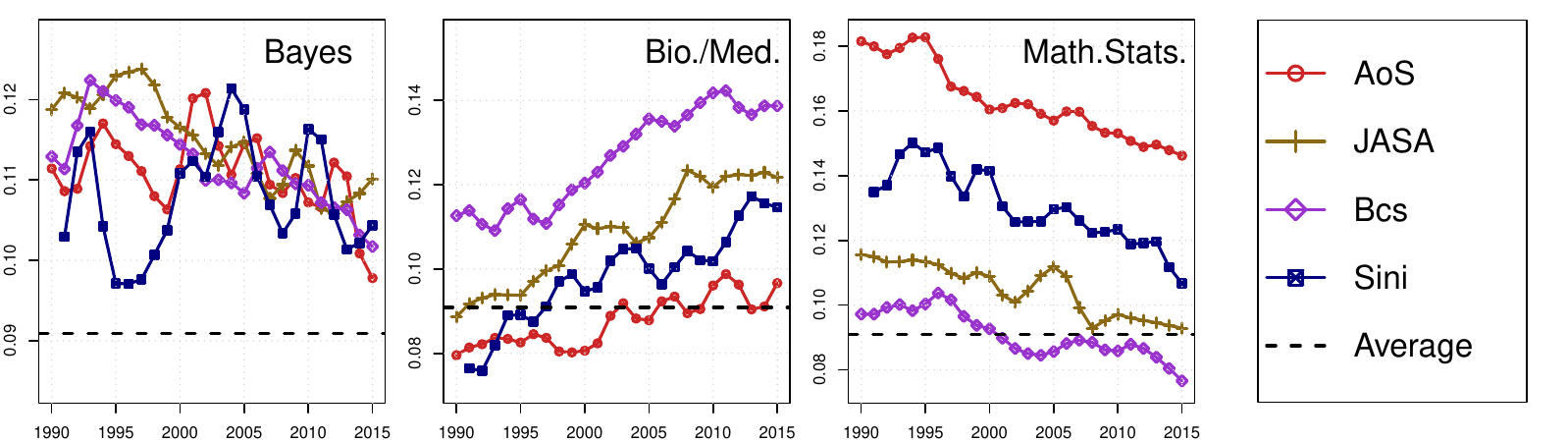}\\
\vspace{.5em}
\scalebox{.72}{
\begin{tabular}{ll | ll | ll | ll}
\hline
Topic  & Journal & Topic & Journal & Topic & Journal & Topic & Journal \\ 
\hline 
Bayes &  Bka & Exp.Design &  Sini & Latent.Var. &  JMLR (04-07)  &   Regression &  AoS (90-02)   \\ 
 &  JRSSB & Hypo.Test & Bcs (90-98) & &  JRSSB (08-12) &  Time Series &  Bcs\\
Bio./Med. &  Bcs  &  & AoS (02-15)  &  Mach.Learn. & JMLR &  \\ 
Clinic. &  Bcs   & Inference &  AoS &  Math.Stats.& AoS    \\
\hline \\ 
\end{tabular}
} 
\caption{\small Top: the yearly average topic weights for selected journals during 1990--2015 (the complete result is in Figure~\ref{fig:tw_dynamic_journal} of supplementary material).
Bottom: the friendliest journal (out of 7 selected journals) for each topic.} \label{fig:tw_dynamic_journal_short}
\end{figure}

Second, we select a few journals and study how the evolution of the yearly average topic weights for each journal. 
In Section~\ref{subsec:topic-JR} we have ranked the $33$ journals (excluding 3 probability journals) by the Stigler's model and PageRank.  
We select the 7 journals with highest average ranks:  {\it AoS}, {\it Bka}, {\it JASA}, {\it JRSSB}, {\it Bcs}, {\it JMLR}, and {\it Sini}. For each journal, we obtain the yearly average topic weight (i.e., the average of $\hat{w}_i$ among papers published in this journal each year) and smooth the curves as before. The results are in Figure~\ref{fig:tw_dynamic_journal} of the supplementary material. A partial result is shown in Figure~\ref{fig:tw_dynamic_journal_short}. Each panel corresponds to a topic. Fixing a topic $k$, for each journal, we plot the $k$th entry (subject to smoothing over time) in the yearly average of $\hat{w}_i$'s among papers published in this journal. These curves of different journals for the same topic can be used to study journal friendliness to this topic.   
 
We observe that in some time periods, some journals are clearly in favor of some topics. When this happens,  we say that this journal is ``friendly" to this topic. 
In Figure~\ref{fig:tw_dynamic_journal_short}, we list the ``friendliest'' journals for 11 topics.   
Note that the short label of a topic may not be accurate for all research topics it covers, and it is preferable to consult Table \ref{tab:11-topic} (e.g. {\it Time Series} includes longitudinal data and survival analysis, and it is why this topic has a high weight in the journal {\it Bcs}).  
Among the 7 journals, {\it JMLR} has a significantly higher weight on {\it Mach.Learn.} than on the other topics,  {\it Bcs} has a significantly higher weight on {\it Bio./Med.} and {\it Clinic.},  and {\it AoS} has a considerably higher weight on {\it Math.Stat.}.  
Furthermore, the 4 journals, {\it AoS}, {\it Bka}, {\it JASA} and {\it JRSSB}, are traditionally considered the leading journals in statistical method and theory. Among these 4 journals, {\it AoS} is friendlier to {\it Math.Stat.}, {\it Inference}, {\it Hypo.Test}, {\it Regression}, and {\it Exp.Design}; {\it JASA} is friendlier to  {\it Mach.Learn.}, {\it Bio./Med.}, {\it Clinic.} and {\it Time Series};  {\it JRSSB} is friendlier to  {\it Mach.Learn.}, {\it Bayes}, and {\it Var.Select.}; and {\it Bka} is friendlier to  {\it Bayes} and {\it Regression} ({\it JASA} publishes more on {\it Clinic.} and {\it Bio./Med.} than {\it Bka};  this is possibly due to that {\it JASA} has a case-study sector).   

\section{TR-SCORE: an extension of Topic-SCORE for topic ranking}
\label{sec:extension}  
Topic-SCORE is a flexible idea and can be extended in many directions. 
In this section,  we extend Topic-SCORE by proposing Topic-Ranking-SCORE (TR-SCORE) 
as new approach to ranking the citation impacts of different topics. 
Since TR-SCORE is directly motivated by the analysis of MADStat, we focus our discussion on the MADStat dataset in this section but keep in mind that the idea is useful in other applications.

In Section \ref{sec:basic}, we have discussed how to use citation exchanges to rank 
different journals. We can extend the idea to topic ranking, but there is a major challenge: 
citation exchanges between papers or journals are well-defined and directly observable, 
but citation exchanges between research topics are not well-defined and directly observable.  
We tackle this by combining the abstracts and the citation data: we first 
propose a model that jointly models text abstracts and citations,  
including an idea to measure the (unobserved) citation exchanges between research topics. 
We then introduce TR-SCORE, and use it to rank different topics and to 
construct a knowledge graph visualizing the cross-topic citation exchanges.

\subsection{The Hofmanm-Stigler model for abstract and citation data} \label{subsec:HofStig}  
Consider $n$ papers in MADStat, where the abstract data are summarized in 
a $p \times n$  word-document-count matrix $X = [x_1, x_2, \ldots, x_n]$ 
as in Section \ref{sec:topic} ($p$ is the vocabulary size), and 
citation data are summarized in an adjacency matrix 
$C \in \mathbb{R}^{n\times n}$,  where $C_{ij} = 1$ if there is a citation from paper $i$ to paper $j$ and $C_{ij} = 0$ otherwise,  $1 \leq i, j \leq n$.  
  
We propose the {\it Hofmann-Stigler model} to jointly model the data matrices $X$ and $C$: It combines the Hofmann's topic model in Section~\ref{sec:topic} and the Stigler's model in Section~\ref{subsec:topic-JR}.    
We assume that all the paper abstracts focus on $K$ different research topics  ${\cal C}_1, {\cal C}_2, \ldots, {\cal C}_K$.  Inspired by the Stigler's model,   
we introduce $\mu=(\mu_1,\mu_2,\ldots,\mu_K)'$, where $\mu_k$ is the {\it export score} associated with 
topic $k$, $1 \leq k \leq K$. Intuitively, a topic with a larger export score means that it has larger impacts.  
Now, fix $1 \leq i \leq n$ and consider paper $i$. 
Similarly as in Section~\ref{sec:topic}, let $w_i \in \mathbb{R}^K$ be 
the weight vector of document $i$ (i.e., $w_i(k)$ is the weight that abstract $i$ puts on topic $k$).   
When paper $i$ is cited by another paper $j$, we have two different ways to attribute this particular citation count.   
\begin{itemize}
\itemsep0em 
\item {\it (Orthodox Citation Attribution (OCA))}.  We simply attribute the citation to paper $i$. 
\item {\it (Topic Weight Citation Attribution (TWCA))}. We attribute the citation to each of the $K$ topics, with weights $w_i(1), \ldots, w_i(K)$, respectively (note that $\sum_{k=1}^K w_i(k) = 1$).   
\end{itemize} 
In Section \ref{subsec:topic-JR}, we have discussed journal ranking, in which OCA is a good choice.  
For topic ranking, TWCA is more appropriate.  
Under TWCA, we view $\mu' w_i  = \sum_{k = 1}^K \mu_k  w_i(k)$ as the 
export score of paper $i$ and assume the Bernoulli variables $C_{ij}$ and $C_{ji}$ satisfy 
\beq \label{Stigler}
\mathbb{P}(C_{ij}=1|C_{ij}+C_{ji}\geq 1) = \frac{\exp(\mu'w_i-\mu'w_j)}{1+\exp(\mu'w_i- \mu'w_j)}. 
\eeq 
This gives the model of the citation exchange matrix $C$. To model the word-document-count matrix $X$, we use the same model as in Section~\ref{sec:topic}:
\begin{equation} \label{Hofmann} 
x_i \sim \mathrm{Multinomial}(N_i, A w_i),  \qquad A \in \mathbb{R}^{p\times K}, \;  w_i \in \mathbb{R}^K,  
\end{equation} 
where $A$ is the topic matrix as in Section~\ref{sec:topic} and $N_i$ is the size (total word count) of document $i$.   
For identifiability, we assume $\mathrm{median}(\mu_1, \ldots, \mu_K) = 0$. 
Also, for simplicity, we assume $X$ and $C$ are independent (but their distributions are related by $w_i$'s), and this can be relaxed. We call \eqref{Stigler}-\eqref{Hofmann} the {\it Hofmann-Stigler} model.


\subsection{Topic-Ranking SCORE (TR-SCORE)} 
We propose TR-SCORE for topic ranking. The input are $X$, $C$, and the number of topics $K$, and the output is an estimated export score vector $\hat{\mu}$. TR-SCORE has three steps.
\begin{enumerate}
\item ({\it Topic matrix estimation}).  Apply Topic-SCORE (e.g., Section~\ref{subsec:topic-SCORE}) to get $\hat{A} \in \mathbb{R}^{p\times K}$. 
\item ({\it Topic weight estimation}).   For $1 \leq i \leq n$, estimate $w_i$ by $\hat{w}_i = (\hat{A}' \hat{A} + \lambda   I_K)^{-1} \hat{A}' d_i$, where $\lambda>0$ is a regularization parameter which we usually fix at $\lambda=0.3$.  
\item ({\it Topic ranking}).  Plug $\hat{w}_1, \ldots, \hat{w}_n$ into (\ref{Stigler}) and obtain an estimate $\hat{\mu}$ for the export score vector $\mu$.  Rank topics according to the descending order of $\hat{\mu}_1,\hat{\mu}_2,\ldots,\hat{\mu}_K$.  
\end{enumerate} 
We discuss Step 3 in detail. We use a quasi-likelihood method with over-dispersion to obtain $\hat{\mu}$. Recall that $C$ is the adjacency matrix of between-paper citations. Write $\bar{C}=C+C'$ (i.e., $\bar{C}_{ij}=C_{ij}+C_{ji}$). Recall that $W=[w_1,w_2,\ldots,w_n]\in\mathbb{R}^{K,n}$ is the topic weight matrix. Let $\tau(x)=e^x/(1+e^x)$ denote the logistic function. We now slightly modify \eqref{Stigler} to assume
\beq \label{quasi-Stigler}
\mathbb{E}[C|\bar{C}]={\bar C}\circ\Omega, \quad \mathrm{Var}(C|\bar{C})=\phi[\Omega\circ(1-\Omega)], \quad\mbox{with}\;\; \Omega=\tau({\bf 1}_n\mu'W-W'\mu{\bf 1}_n'),
\eeq 
where $\circ$ is the Hadamard product, $\mathrm{Var}(C|\bar{C})$ and $(1-\Omega)$ are both element-wise operations, and $\phi>0$ is the dispersion parameter. Model~\eqref{Stigler} corresponds to fixing $\phi=1$, but a better strategy is to estimate $\phi$ from data, as commonly used in fitting count data (e.g., see \cite{Va2016} for a similar strategy for fitting the Stigler's model). When $W$ is known, we estimate $\mu_1,\mu_2,\ldots,\mu_K$ by maximizing the quasi-likelihood, 
which is equivalent to maximizing the likelihood of model \eqref{Stigler}. This is done by first fixing $\mu_1=0$ and treating \eqref{Stigler} as a generalized linear model with $(K-1)$ predictors and $N:=\sum_{i,j}1\{\bar{C}_{ij}= 1\}$ samples, so that it can be solved by a standard package. We then re-center $\hat{\mu}_1,\hat{\mu}_2,\ldots,\hat{\mu}_K$ so that their median is $0$. The dispersion parameter is estimated by $
\hat{\phi} = \frac{1}{N-K+1}\sum_{(i,j): i<j, \bar{C}_{ij}\geq 1} (C_{ij}-\bar{C}_{ij}\hat{\Omega}_{ij})^2/[\bar{C}_{ij}\hat{\Omega}_{ij}(1-\hat{\Omega}_{ij})]$, 
where $\hat{\Omega}_{ij}=\tau(\hat{\mu}'w_i-\hat{\mu}'w_j)$. So far, $W$ is assumed known. For unknown $W$, we use the same procedure, except that $W$ is replaced by the $\hat{W}$ from Step 2.

\subsection{Topic-ranking and a cross-citation graph} \label{subsec:citation-graph}
In Section~\ref{sec:application}, we have applied 
Topic-SCORE to a set of 56,500 (pre-processed) abstracts and identified 11 
representative research topics in statistics. 
We now use TR-SCORE to the same set of abstracts and rank all 11 topics. 
We also build a cross-topic citation graph (as a type of knowledge graph) to visualize the dissemination of knowledge across areas (an important research topic in the area of modern knowledge discovery \citep{Shi2015weaving}).

\begin{figure}[tb!]
	\centering
        \includegraphics[height=.4\textwidth]{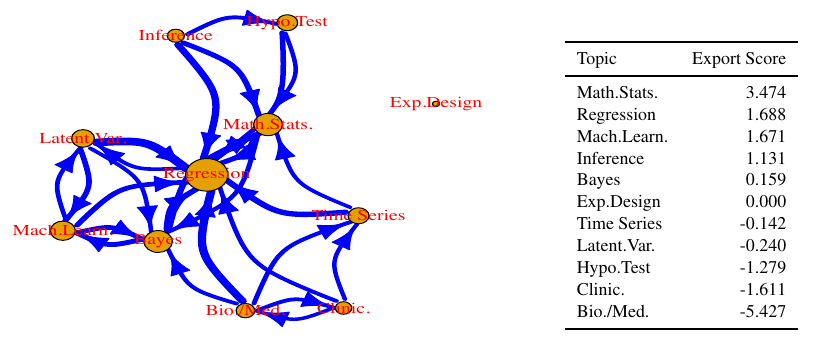}
	\caption{\small Left: The weighted directed graph for cross-topic citations. The diameter of a node (topic) is proportional to the total citations the topic has received from other topics, and the width of an edge is proportional to the weight defined in the text.  An edge is presented if the weight is bigger than $0.09$. Right: The estimated export scores of 11 topics (subject to $\mathrm{median}(\hat{\mu}_1,\ldots,\hat{\mu}_{11})=0$).}  \label{fig:TopicCitations}
\end{figure}

We first build a cross-topic citation graph. This is a weighted and directed graph with 11 nodes, each being a discovered topic. We propose two definitions of edge weights. 
In the first one, let $N_{k, \ell} = \sum_{i, j =1 }^n \hat{w}_i(k) \hat{w}_j(\ell) C_{ij}$ and $P_{k\ell}=N_{k\ell}/(\sum_{m=1}^KN_{km})$, for $1 \leq k,\ell \leq 11$,  where $C$ is the between-paper citation adjacency matrix and $\hat{w}_i$ is the topic weight vector of abstract $i$. Here $N_{k\ell}$ is the (allocated) citation counts from topic $k$ to topic $\ell$, and $P_{k\ell}$ is the proportion of citations to topic $\ell$ among all citations from topic $k$. We use $P\in\mathbb{R}^{11\times 11}$ as the weighted adjacency matrix of this graph. In the second definition, we group all papers based on the `dominant topic' - the topic with the largest weight in $\hat{w}_i$ (if there is a tie, pick the smaller $k$). Let $w_i^*\in\{e_1,e_2,\ldots,e_K\}$ denote the group label of abstract $i$. Define $N^*_{k, \ell} = \sum_{i, j =1 }^n \hat{w}^*_i(k) \hat{w}^*_j(\ell) C_{ij}$ and $P^*_{k\ell}=N^*_{k\ell}/(\sum_{m=1}^KN^*_{km})$. We then use $P^*\in\mathbb{R}^{11\times 11}$ as the weighted adjacency matrix.  This definition uses ``winner takes all'' to allocate each citation to a single pair of topics.    
The two matrices $P$ and $P^*$ are shown in Tables~\ref{tb:cross-topic-citation}-\ref{tb:cross-topic-citation-2} of the supplementary material. 
Both definitions make sense, but the second one leads to a `sparser' graph, which is presented in Figure~\ref{fig:TopicCitations} (the first one is relegated to Figure~\ref{fig:TopicCitations-2}).  

In Figure~\ref{fig:TopicCitations} (left), the width of the edge from node $k$ to node $\ell$ is proportional to $P^*_{k\ell}$, and the edge is presented only when $P^*_{k\ell} \geq 0.09$.   We have interesting observations.   First, {\it Exp.Design} has relatively few citation exchanges with other topics and the majority of the citations it receives are from the topic itself. Since a one-way edge from node $k$ to node $\ell$ is presented when $P^*_{k\ell} \geq 0.09$,  no edge from or to {\it Exp.Design} is shown in Figure \ref{fig:TopicCitations}. 
Second, {\it Regression} and {\it Math.Stat.} are the two topics that have  attracted the most citations from  other topics, and  {\it Bio./Med.} and {\it Inference} are the two  that have  cited 
other topics most often.  Third, each of {\it Bayes}, {\it Variable Selection}, and {\it Mach.Learn.} has 
considerably many outgoing and incoming citations. 
Last,  {\it Hypo.Test} and {\it Inference} form a close pair, and most in-between citations are from {\it Inference} to {\it Hypo.Test}; {\it Clinic.} and {\it Bio./Med.} form a close pair, and the citation exchanges are relatively balanced between them.

We then consider topic ranking. Figure~\ref{fig:TopicCitations} (right) shows the export scores of 11 topics by TR-SCORE.  
{\it Math.Stats.} is the highest-ranked topic. This is reasonable, as the focus of {\it Math.Stats.} is mathematical analysis and probability, which may have a long-lasting impact on other topics in statistics. {\it Regression} and {\it Mach.Learning} are 
also highly ranked. This is also understandable, as the two topics cover many ``hot'' research topics (see Table~\ref{tab:11-topic}). The rankings of {\it Bio./Med.} and {\it Clinic.} are relatively low; one reason is that a significant fraction of their impacts are over research areas outside our data range.

\section{Conclusion} \label{sec:conclusion}

Text analysis is a rapidly developing research area in data science. In this paper, we have surveyed recent methods for text analysis, ranging from topic modeling to neural language models.
For topic modeling, we have discussed the anchor word condition, several different algorithms, optimal rates, and extensions to bigram and supervised models. In particular, we focus on Topic-SCORE, a fast algorithm that enjoys appealing theoretical properties. 
For neural language models, we provided a brief introduction to its key components, reviewed the popular BERT and word embedding models, and discussed how to apply them to solve downstream NLP tasks.  

We have also presented a data set, MADStat, about academic publications in statistics. It was collected and cleaned by ourselves with substantial efforts. We have made it publicly available at \url{http://zke.fas.harvard.edu/MADStat.html}. In this paper, we analyzed text abstracts of the papers in MADStat, using the Topic-SCORE algorithm. We discovered 11 representative topics and visualized the trends and pattens in statistical research. We also proposed the Hoffman-Stigler model to jointly model text abstracts and citation data and the TR-SCORE algorithm for ranking the citation impacts of 11 topics. These results are not only applications of text analysis but also can be viewed as a data-driven review of the academic statistical community. 

Nowadays, a vast amount of text data are generated on a daily basis. Recent advancements in Natural Language Processing (NLP) have revolutionized our everyday lives. This also provides a big opportunity to statistics. 
The statistical approaches to NLP are typically transparent, sample-efficient, fast-to-compute, and theoretically tractable, making them a suitable choice for many ordinary NLP users (who may have a moderate-size domain-specific corpus but cannot access the data and resources owned by those tech giants). 
On the other hand, statistical text analysis is still quite under-developed. 
Even for topic modeling, there are still many unresolved problems, such as how to estimate the number of topics and how to improve the accuracy when the documents are extremely short. 
We hope that this review article provides useful information to researchers interested in this area. We also hope that the MADStat dataset, which we collected and shared with public, serves as a good platform for testing existed methods and inspiring new research in text analysis.

\appendix

\newpage 
\section{Data collection and cleaning} \label{sec:data-collection}
One might think that our data sets is easy to obtain, as it seems that BibTeX and citation data are easy to download.  Unfortunately, when we need a large-volume high-quality data set, this is not the case. For example, the citation data by Google Scholar is not very accurate, and many online resources do not allow for large volume downloads.  Our data are downloaded using a handful of techniques including, but not limited to, web scraping. The data set was also carefully cleaned by a combination of manual efforts and computer algorithms we developed. Both data collection and cleaning are sophisticated and time-consuming processes, during which we have encountered a number of challenges.

The first challenge is that, for many papers, we need multiple online resources to acquire the complete information.  For example, to download complete information of a paper, we might need online resources 1, 3, and 5 for paper 1, whereas   online resources 2, 4, and 6  for paper 2. Also, each online resource may have a different system to label their papers. As a result, we also need to carefully match papers in one online resource to the same ones in another online resource. These make the downloading process rather complicated.

The second challenge is name matching and cleaning. For example, some journals list the authors only with the last name and first initial, so it is hard to tell whether “D. Rubin” is Donald Rubin or Daniel Rubin. Also,  the name of the same author may be spelled differently in different papers (e.g.,  “Kung-Yee Liang”  and  “Kung Yee Liang”).    A more difficult case is that different authors may share the same name (e.g., Hao Zhang at Purdue University and Hao Zhang at Arizona State University). To correctly match the names and authors, we have to combine manual efforts with some computer algorithms.

Last, an online resource frequently has internal inconsistencies, syntax errors, encoding issues, etc. We need a substantial amount of time and efforts to fix these issues.

\section{Disclaimer} \label{sec:disclaimer}
It is not our intention to rank a researcher (or a paper, or an area) over others. For example, when we say a paper is “highly cited,” we only mean that the citation counts are high, and we do not intend to judge how important or influential the paper is. Our results on journal ranking are based on journal citation exchanges, but we do not intend to interpret the ranking more than the numerical results we obtain from the algorithms we use.

As our data set is drawn from real-world publications, we have to use real names, but we have not used any information that is not publicly available. For interpretation purposes, we frequently need to suggest a label for a research group or a research area, and we wish to clarify that the labels do not always accurately reflect all the authors/papers in the group. Our primary interest is the statistics community as a whole, and it is not our intention to label a particular author (or paper, or topic) as belonging to a certain community (group, area).

While we try very hard to create a large-scale and high-quality data set, the time and effort one can invest in a project is limited. As a result, the scope of our data set is limited. Our data set focuses on the development of statistical methods and theory in the past 40 years, and covers research papers in 36 journals between 1975 and 2015 (we began downloading data in 2015). These journals were selected from the 175 journals on the {\it 2010 ranked list of statistics journals by the Australian Research Council} (see Section~\ref{sec:journal-list}).  Journals on special themes and most journals on econometrics, interdisciplinary research, and applications are not included (see Section 6.1 for detailed description). As a result, papers on econometrics, interdisciplinary research, and applications may be underrepresented.

Due to the limited scope of our data set, some of our results may be biased. For example, for the citations a paper has received, we count only those within our data range, so the resultant citation counts may be lower than the real counts the paper has received. Alternatively, for each paper, we can count the citation by web searching (e.g., Google Scholar, which is known to be not very accurate), or by reference matching (e.g., Web of Science and Scopus). Our approach allows us to perform advanced analysis (e.g., ranking authors/papers by citation counts, reporting the most cited authors and papers, excluding self-citations, and calculating cross-journal citation). For such analysis, it is crucial that we know the title, author, author affiliation, references, and time and place where it is published for each paper under consideration. For each of the two alternative approaches, we can gather such information for a small number of papers, but it is hard to obtain such information for 83,336 papers as in our data set.

A full scope study of a scientific community is impossible to accomplish in one paper. The primary goal of our paper is to serve as a starting point for this ambitious task by creating a template where researchers in other fields (e.g., physics) can use statisticians' expertise in data analysis to study their fields.  
For these reasons, the main contributions of our paper are still valid, 
despite some limitations discussed above.

\spacingset{1.1}
\begin{table}[bt!]
	\centering
		\caption{For each of the 36 journals, we present the full name, abbreviated name, starting time, total number of authors, total number of papers, and impact factors 
		in 2014 and 2015. For each journal, our data set consists of all papers between a certain year (i.e., the starting time) and 2015. The starting time is not necessarily the year the journal was launched.} 
	\label{tab:journal}%
	\scalebox{0.8}
	{
		\begin{tabular}{|r|l|l|c|c|c|c|c|}
			\hline
			& & {\small Abbrev.} & {\small Starting} & {\small  $\#$ of}  & {\small  $\#$ of } &   &  \\  
			& Full name of the journal & {\small Name} & {\small Time} & {\small Papers} & {\small Authors} & {\small IF2014} & {\small IF2015} \\
			\hline
			1      & \textit{Ann. Inst. Henri Poincare Probab. Stat.} & AIHPP  & 1984   & 967    & 1152   & 1.27   & 1.099 \\
			2      & \textit{Annals of Applied Statistics} & AoAS   & 2007   & 729    & 1824   & 0.942  & 0.769 \\
			3      & \textit{Annals of Probability} & AoP    & 1975   & 3318   & 2277   & 2.032  & 1.842 \\
			4      & \textit{Annals of Statistics} & AoS    & 1975   & 4168   & 3065   & 1.729  & 1.968 \\
			5      & \textit{Annals of the Institute of Statistical Mathematics} & AISM   & 1975   & 2016   & 2056   & 3.055  & 3.528 \\
			6      & \textit{Australian \& New Zealand Journal of Statistics} & AuNZ   & 1998   & 592    & 968    & 0.509  & 0.62 \\
			7      & \textit{Bayesian Analysis} & Bay    & 2006   & 138    & 314    & 1.519  & 1.031 \\
			8      & \textit{Bernoulli} & Bern   & 1997   & 1065   & 1446   & 1.829  & 1.412 \\
			9      & \textit{Biometrics} & Bcs    & 1975   & 4347   & 5357   & 1.491  & 1.603 \\
			10     & \textit{Biometrika} & Bka    & 1975   & 3359   & 3239   & 2.94   & 2.114 \\
			11     & \textit{Biostatistics} & Biost  & 2002   & 732     & 1575    & 1.642  & 1.336 \\
			12     & \textit{Canadian Journal of Statistics} & CanJS  & 1985   & 1202   & 1542   & 1.676  & 1.41 \\
			13     & \textit{Communications in Statistics-Theory and Methods} & CSTM   & 1976   & 8390   & 8041   & 0.424  & 0.437 \\
			14     & \textit{Computational Statistics \& Data Analysis} & CSDA   & 1983   & 4656   & 6725   & 0.713  & 0.6 \\
			15     & \textit{Electronic Journal of Statistics} & EJS    & 2007   & 703    & 1156   & 1.303  & 0.903 \\
			16     & \textit{Extremes} & Extrem & 2008   & 176    & 262    & 1.5    & 1.68 \\
			17     & \textit{International Statistical Review} & ISRe   & 1975   & 855    & 1128   & 2.081  & 1.711 \\
			18     & \textit{Journal of Computational and Graphical Statistics} & JCGS   & 1997   & 907    & 1488   & 2.319  & 2.038 \\
			19     & \textit{Journal of Machine Learning Research} & JMLR   & 2001   & 1332   & 2362   & 1.544  & 2 \\
			20     & \textit{Journal of the American Statistical Association} & JASA   & 1975   & 5154   & 5686   & 0.939  & 1.676 \\
			21     & \textit{Journal of the Royal Statistical Society} & JRSSB  & 1975   & 1682   & 1882   & 2.742  & 3.125 \\
			& \textit{Series B-Statistical Methodology} &   &    &     &   &      &  \\
			22     & \textit{Journal of Applied Statistics} & JoAS   & 1993   & 2219   & 3798   & 1.18   & 1.058 \\
			23     & \textit{Journal of Classification} & JClas  & 1984   & 435    & 551    & 0.569  & 0.587 \\
			24     & \textit{Journal of Multivariate Analysis} & JMVA   & 1976   & 3574   & 3601   & 2.286  & 2.357 \\
			25     & \textit{Journal of the Royal Statistical Society} & JRSSA  & 1975   & 1117    & 1821   & 4      & 5.197 \\
			& \textit{Series A-Statistics in Society} &   &    &     &   &      &  \\
			26     & \textit{Journal of the Royal Statistical Society} & JRSSC  & 1975   & 1359    & 2282   & 1.753  & 1.615 \\
			& \textit{Series C-Applied Statistics} &   &    &     &   &      &  \\
			27     & \textit{Journal of Statistical Planning and Inference} & JSPI   & 1977   & 6111   & 6372   & 0.818  & 0.869 \\
			28     & \textit{Journal of Time Series Analysis} & JTSA   & 2000   & 692    & 925    & 0.939  & 1.387 \\
			29     & \textit{Journal of Nonparametric Statistics} & JNS    & 1998   & 817    & 1187   & 0.586  & 0.556 \\
			30     & \textit{Probability Theory and Related Fields} & PTRF   & 1986   & 2164   & 1874   & 1.657  & 2.025 \\
			31     & \textit{Statistical Science} & StSci  & 1993   & 564    & 980    & 1.59   & 1.641 \\
			32     & \textit{Scandinavian Journal of Statistics} & ScaJS  & 1977   & 1393   & 1730   & 2.154  & 1.741 \\
			33     & \textit{Statistica Sinica} & Sini   & 1991   & 1685   & 2235   & 0.718  & 0.63 \\
			34     & \textit{Statistics and Computing} & SCmp   & 1993   & 907    & 1518   & 1.032  & 1.155 \\
			35     & \textit{Statistics \& Probability Letters} & SPLet  & 1984   & 7063   & 6670   & 1.382  & 0.952 \\
			36     & \textit{Statistics in Medicine} & SMed   & 1984   & 6743   & 9575   & 2.942  & 2.817 \\
			\hline
		\end{tabular}%
	}
\end{table}%

\spacingset{1.475}

\section{The list of 36 journals}\label{sec:journal-list}

The $36$ journals are selected as follows.  We start with the $175$ journals in the 2010 ranked list of statistics journals provided by the Australian Research Council (ARC). \footnote{\url{https://www.righttoknow.org.au/request/616/response/2048/attach/3/2010}.}   The list was used for performance evaluation of Australian universities, as part of its program of {\it Excellence in Research for Australia}. 
The 175 journals are divided into four categories:  $A^*$, $A$,  $B$, and $C$.   
For our study, first, we include all 9 Category $A^*$ journals,  where  two of them  (AOP and PTRF) are probability journals.  Second, we include all Category $A$ journals, except the strongly themed journals in applied probability or in engineering (Advances in Applied Probability, Electronic Journal of Probability, Finance and Stochastics, Journal of Applied Probability, Stochastic Processes and their Applications, Theory of Probability and its Applications, Technometrics, Queueing Systems, Random Structures $\&$ Algorithms).  Last,  there are about $50$ journals in  Category B covering a wide range of themes, where we only select the journals on methodology and theory, such as  Australian $\&$ New Zealand Journal of Statistics, Bayesian Analysis, Canadian Journal of Statistics, etc.   We do not include any Category $C$ journals.

\section{Additional results on paper counts} \label{sec:paper-counts} 

\begin{figure}[htb!]
		\centering  
		\includegraphics[height=0.24\textwidth, width = 0.36\textwidth]{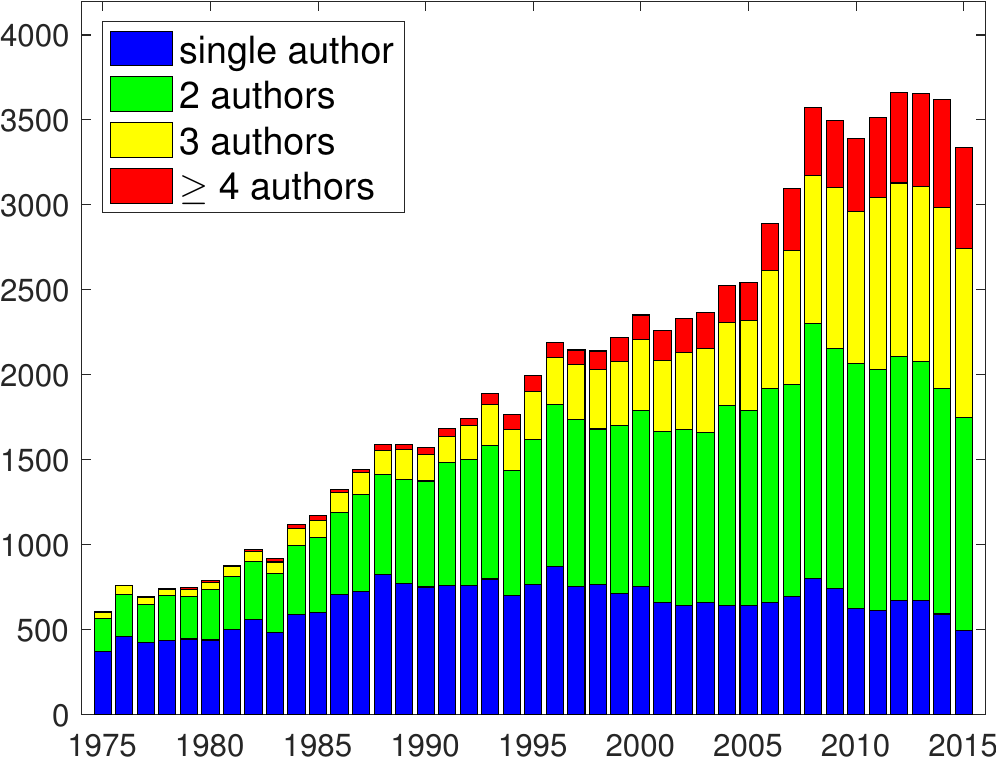} \hspace{1em}  
		\includegraphics[height=0.24\textwidth, width = 0.36\textwidth]{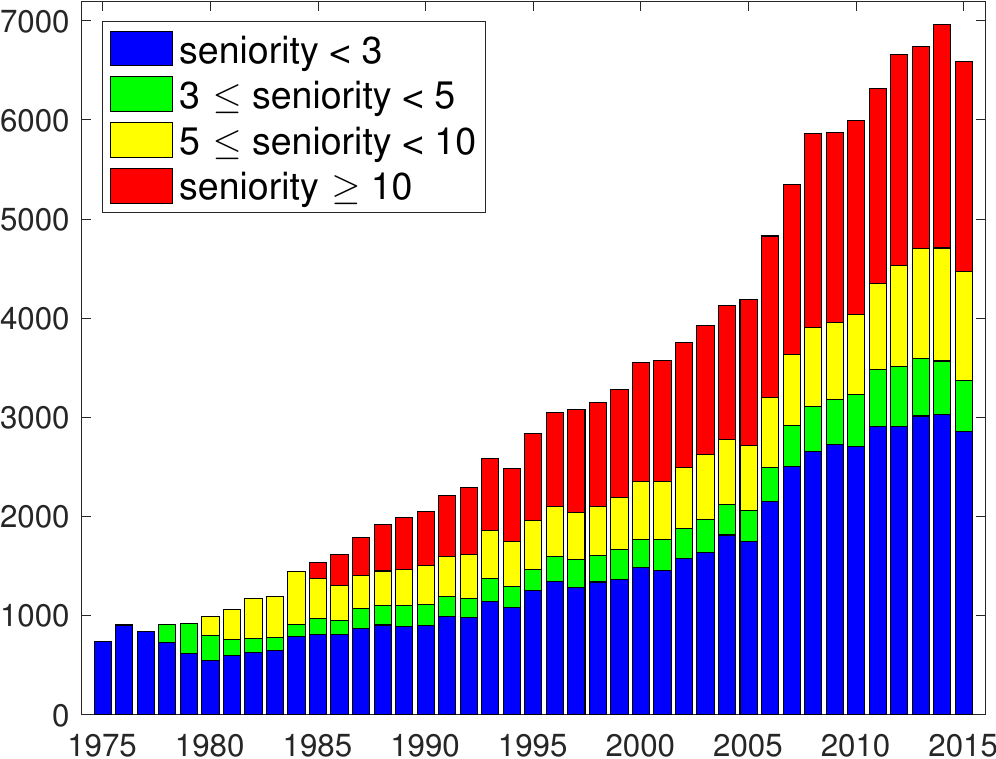}
		\caption{Left: the total number of $m$-authored papers in each year, for different $m$. Right: the total number of $k$-year-senior active authors in each year, for different $k$ (an author who publishes the first paper in year $t$ has a seniority of $k$ in year $t+k$).} 
		\label{fig:productivity2} 
\end{figure}

Figure \ref{fig:productivity1}  presents the 
number of papers per year and the number of  {\it active} authors per year. These results can be further explained using  Figure~\ref{fig:productivity2}.
In Figure~\ref{fig:productivity2} (left), we present the number of $m$-authored papers in each year for $m = 1, 2, 3$ and $m \geq 4$, respectively. It is seen that the fraction of single author papers 
have been steadily decreasing, and the fraction of papers with $3$ or more authors 
have been steadily increasing.  
One possible reason is that, as statistics becomes increasingly more interdisciplinary, publishing in statistical  journals has been increasingly more challenging, as statisticians need to coauthor with researchers from other scientific areas, for their data sets or expertise in their areas, and often works on methods and theory alone are not adequate for publication. Figure~\ref{fig:productivity2} (right)  presents the number of active authors with $k$-year seniority in each year for $k$ in some different ranges.  We say that an author is $k$-year-senior  in  year $t$ if this author's first paper appears in year $t-k$ in our data set.  The plot shows a significant increase of authors with seniority $<$ 3 years, suggesting that the statistics community has attracted more and more junior authors. The cohort with seniority $<$ 3 years and the cohort with seniority $>$ 10 years have the largest and second largest fractions. One possible explanation is that a more senior author tends to have more junior collaborators (e.g., a senior professor tends to have more Ph.D students than a less senior professor); such forged collaborations have improved the productivity 
of both the senior cohort and the junior cohort.  
\spacingset{1.1}
%
\spacingset{1.475}

\section{Additional results on network centrality} \label{sec:center} 
%
%
Table~\ref{tab:toppapers} presents the top 10 most-cited papers. Note that the numbers of coauthors, citers, and citations here are all counted using {\it only} the papers in our data range, so there may be some biases in our ranking.  See Section \ref{sec:network} 
for more discussion on ranking (and especially 
comparisons between ranking with our data set and ranking with the Google Scholar data).   

\spacingset{1.1}
\begin{table}[htb!]
\centering
	\vspace{.5em}
	\caption{The most-cited papers (only the citations within MADStat are counted).} 
	\label{tab:toppapers}  
	\scalebox{0.75}{ 
		\begin{tabular}{llllll} 
			Rank & Author                                                                   & Year & Title                                                                                          & Journal & Citation \\ \hline
			1    & Dempster {\it et al}.                          & 1977 & Maximum likelihood from incomplete data via EM algorithm                                       & JRSSB   & 2241     \\
			2    & Liang \& Zeger                                             & 1986 & Longitudinal data-analysis using generalized linear-models                                     & Bka     & 1437     \\
			3    & Tibshirani & 1996 & Regression shrinkage and selection via the Lasso                                               & JRSSB   & 1327     \\
			4    & Gelfand \& Smith                                              & 1990 & Sampling-based approaches to calculating marginal densities                                    & JASA    & 950      \\
			5    & Laird \& Ware                                                   & 1982 & Random-effects models for longitudinal data                                                    & Bcs     & 844      \\
			6    & Rubin & 1976 & Inference and missing data                       & Bka     & 834      \\
			7    & Efron & 1979 & Bootstrap methods - another look at the Jackknife              & AoS     & 789      \\
			8    & Fan \& Li & 2001 & Variable selection via  nonconvex  $\ldots$ and its oracle properties             & JASA    & 775      \\
			9    & Benjamini \& Hochberg   & 1995 & Controlling the False Discovery Rate - A  $\ldots$  multiple testing & JRSSB   & 695      \\
			10   & Breslow \& Clayton & 1993 & Approximate inference in generalized linear mixed models                                       & JASA    & 689      \\
		\end{tabular}
	} 
\end{table}
\spacingset{1.475}

\section{The sleeping beauty citation patterns}

The ``sleeping beauty" pattern is especially interesting.  To identify papers with such a pattern, 
we need a metric. We adapt the approach in \cite{ke2015defining}.  Fix a paper $i$.  Suppose $T_i$ years (or months/quarters) have passed since its publication by the end of 2015. 
Let $n_i(t)$, $1 \leq t \leq T_i$,  be the number of citations the paper receives in year $t$. Suppose the citation counts reach the peak at 
year $t = t_i^*$. The sleeping beauty metric is defined to be
\begin{equation} \label{eq:sleepingbeauty}   
B_i  =  \sum_{t: 1\leq t \leq t_i^*}   \bigl[n_i(t_i^*)/t_i^* - n_i(t)/t\bigr] / [(n_i(t) \vee 1)/t] .   
\end{equation} 
Intuitively, between Year $1$ and $t_0$, the citation counts may  grow superlinearly,  linearly, or sublinearly, and  $B_i$ is positive, approximately $0$, or negative, respectively.  If  paper $i$ is 
a sleeping beauty, then we expect that (a) $n_i(t_i^*)$ (maximum number of yearly citations) is large, and (b)  
$B_i$ is large (i.e., we expect the citation counts to grow superlinearly between Year $1$ and $t_i^*$ so $B_i$ is large).  Note also that for a sleeping beauty, the citation counts may drop after Year $t_i^*$ but should remain at a relatively high level for at least a few more years.


Since ``sleeping beauty'' is a special kind of highly cited papers,   we start by selecting the 300 papers with the largest maximum number of yearly citations. We then arrange all papers according to the sleeping beauty measure $B_i$. 
Table \ref{tab:sb} presents the $14$ papers (among the 300) with the largest $B_i$, 
and Figure \ref{fig:sb} of the supplement  presents the citation curve $n_i(t)$ for the first $8$ papers on the list.  
All of these papers show a clear sleeping beauty  pattern, suggesting that the introduced measure is reasonable.

\spacingset{1.1}
\begin{table}[htb!]
\centering
	\caption{The $14$ papers with the largest sleeping beauty measures $B$ (among the 300 papers that have the largest maximum yearly citation counts).  TC is total citation counts.}
	\label{tab:sb} 
	\scalebox{0.85}{ 
		\begin{tabular}{llll||llll} 
			Paper       &    Journal & $TC$ & $B$  &   Paper & Journal & TC & $B$  \\ 
			\hline 
			1. Tibshirani (1996)  & JRSSB & 1327 & 145 & 8. \hspace{0.01 em}   Bai \& Saranadasa (1996)  &Sini & 86 & 77 \\
			2. Azzalini (1985)  & ScaJS & 288 & 139 & 9. \hspace{0.01 em}  Holm (1979)  & ScaJS & 265 & 75 \\
			3. Hubert \& Arabie (1985)  & JClas & 179 & 115 & 10. Clayton (1978)  & Bka & 393 & 70 \\
			4. Hill (1975)  & AoS & 280 & 82  & 11.   Fan \& Li (2001)   & JASA & 775 & 69 \\ 
			5. Marcus et al. (1976)   &  Bka & 218 & 80 & 12.  Turnbull (1976)  & JRSSB & 346  & 69\\
			6. Lunn et al. (2000)   & SCmp & 198  & 79 & 13. Pickands (1975)   & AoS & 234 & 67 \\
			7. Rosenbaum \& Rubin (1983)  &Bka & 413 & 78 &  14. Benjamini \& Hochberg  (1995)   & JRSSB & 695 &65 \\
			\hline 
		\end{tabular}
	}
\end{table}
\spacingset{1.475}


\spacingset{1.1}
\begin{figure}[htb!]
	\centering
	\includegraphics[width=0.85\textwidth]{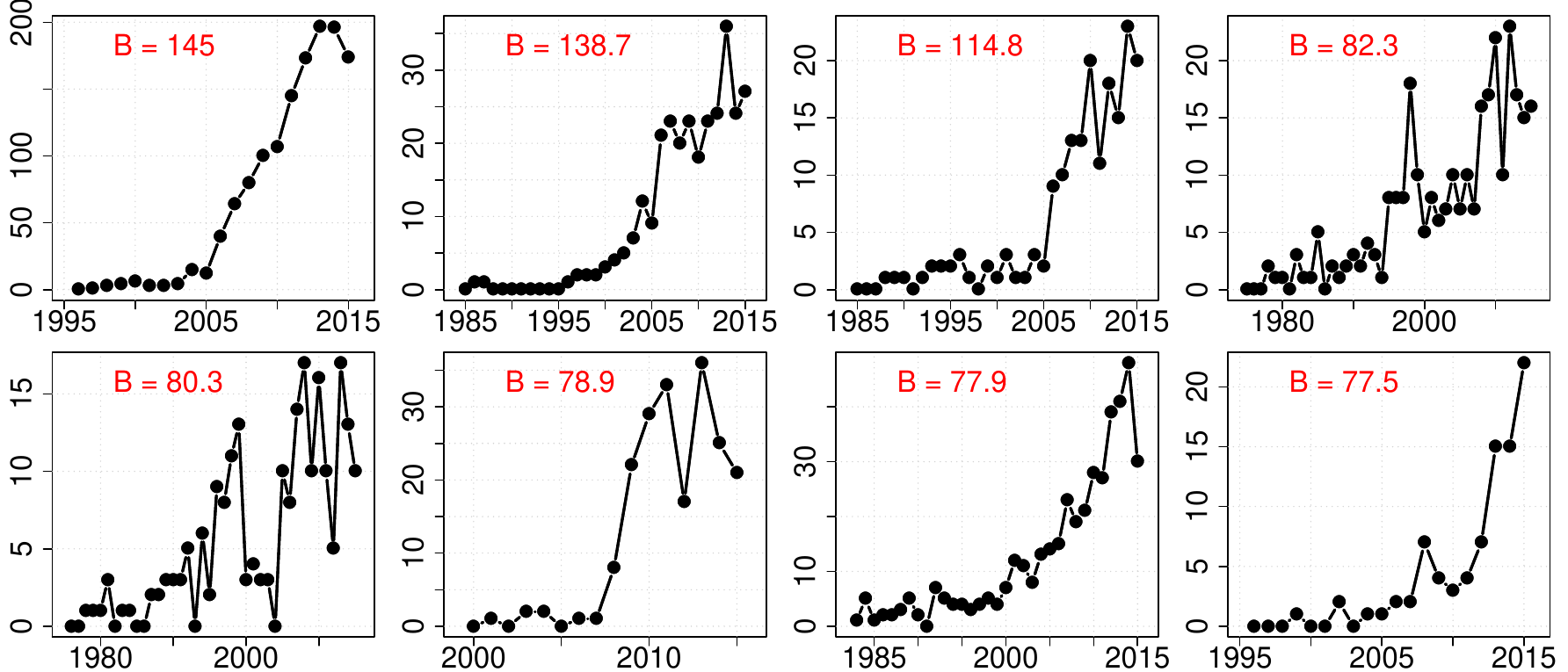}
	\caption{The yearly citation curves for the first $8$ papers in Table \ref{tab:sb}.}
	\label{fig:sb}
\end{figure}
\spacingset{1.475}

\section{Pre-processing of the abstract data} \label{sec:text-preprocessing}


The standard preprocessing includes: 
(i) tokenization, which breaks each abstract into a bag of words; (ii) removing numbers and punctuations; (iii) removing stop words, such as {\it a}, {\it the}, {\it this}, {\it those}, {\it me}; 
and (iv) stemming, which helps unify different forms of the same word, such as {\it testing}, {\it test}, and {\it tests}.
The default functions in the R package \texttt{tm} are not customized for the content of statistical abstracts. We thus add some manual adjustment.

First, our dictionary only allows single words, and for important phrases we must include, 
we have to suppress them first.  
For example, when tokenizing the documents, we encounter phrases such as  {\it test error} and {\it monte carlo}. We suppress them by  {\it testerror} and 
 {\it montecarlo} respectively,  before we insert them to the dictionary.  Second, stemming may sometimes mistakenly combine words with significantly different meanings. For example,  the words {\it measurement} and {\it measure} have the same stem {\it measur}, but very different meaning in our context.  To make sure 
that they are stemmed differently, we replace {\it measurement} by {\it measurement1} before stemming, 
so the stems of {\it measurement} and {\it measure} become {\it measur1} and {\it measur} respectively.  
  Third, the default stop word list in the R package \texttt{tm} does not cover all ``topic-irrelevant words" for the analysis of statistical abstracts. We manually add a list of 289 words (some overlap with the default stop words) to the stop word list. These words include (a) common words used in statistical abstracts, such as {\it data}, {\it estimation}, {\it paper}, {\it method}, {\it propose}, and {\it discuss}; (b) 
words related to the copyright information of the journal or the press, such as 
{\it springer}, {\it wiley}, {\it royal}, and {\it sinica}; and (c) words arising from citing references in the abstract, such as {\it bickel}, {\it berger}, and {\it fan}. 

After the above steps, the vocabulary contains more than $60,000$ words, the majority of which have extremely low frequencies in the corpus. Additionally, some abstracts become quite short after removing stop words. As argued in \cite{KaW2017}, removing low-frequency words and short documents can increase the signal-to-noise ratio. To this end, we first remove all words that appear in fewer than 100 abstracts. This reduces the vocabulary to $p=2,106$. We then remove approximately the 10\% shortest abstracts and retain a total of $n=56,500$ abstracts.

\section{Selection of the number of topics $K$} \label{sec:SelectK}
\begin{figure}[htb]
\centering
\includegraphics[width = 0.4\textwidth, height=0.26\textwidth,  trim=20 40 20 50, clip=true]{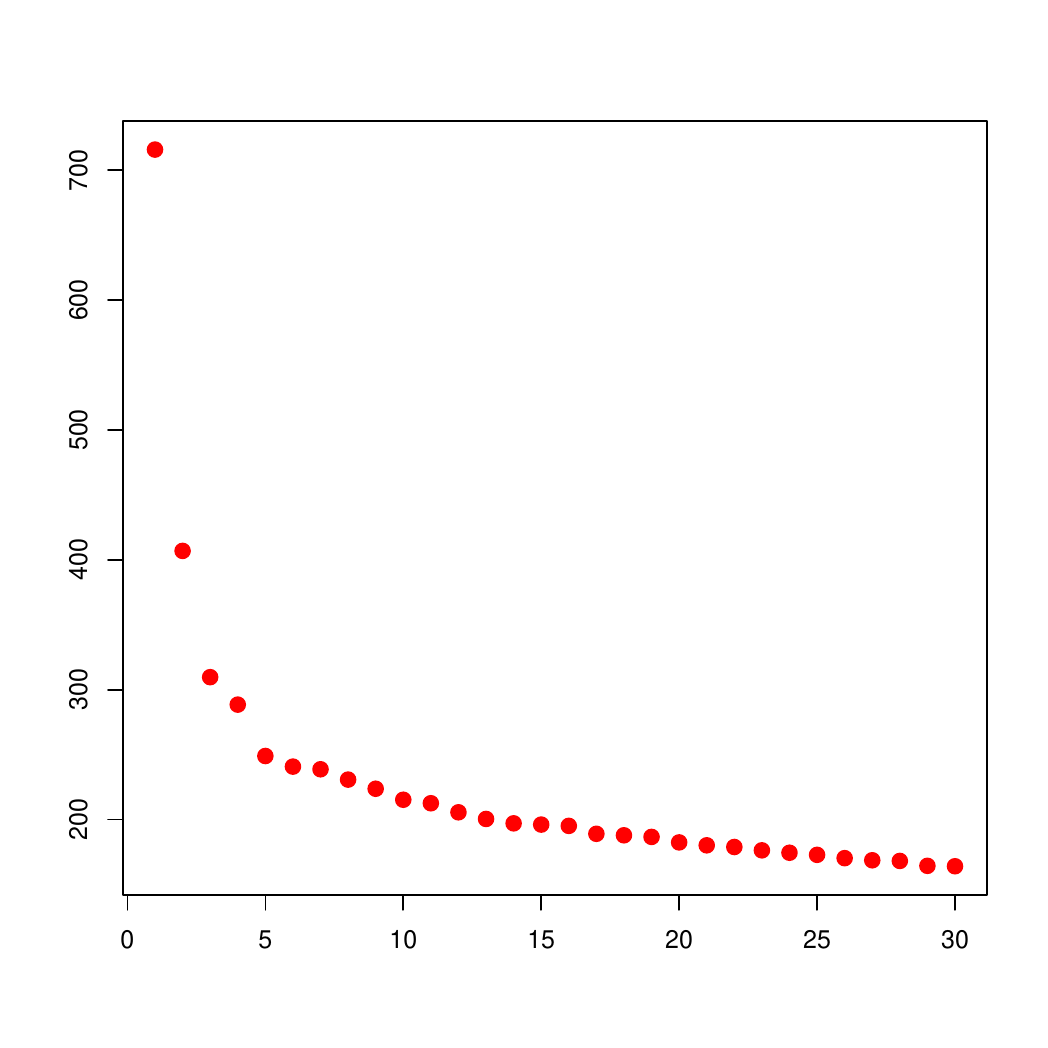}
\includegraphics[width = 0.4\textwidth, height=0.26\textwidth, trim=20 40 20 50, clip=true]{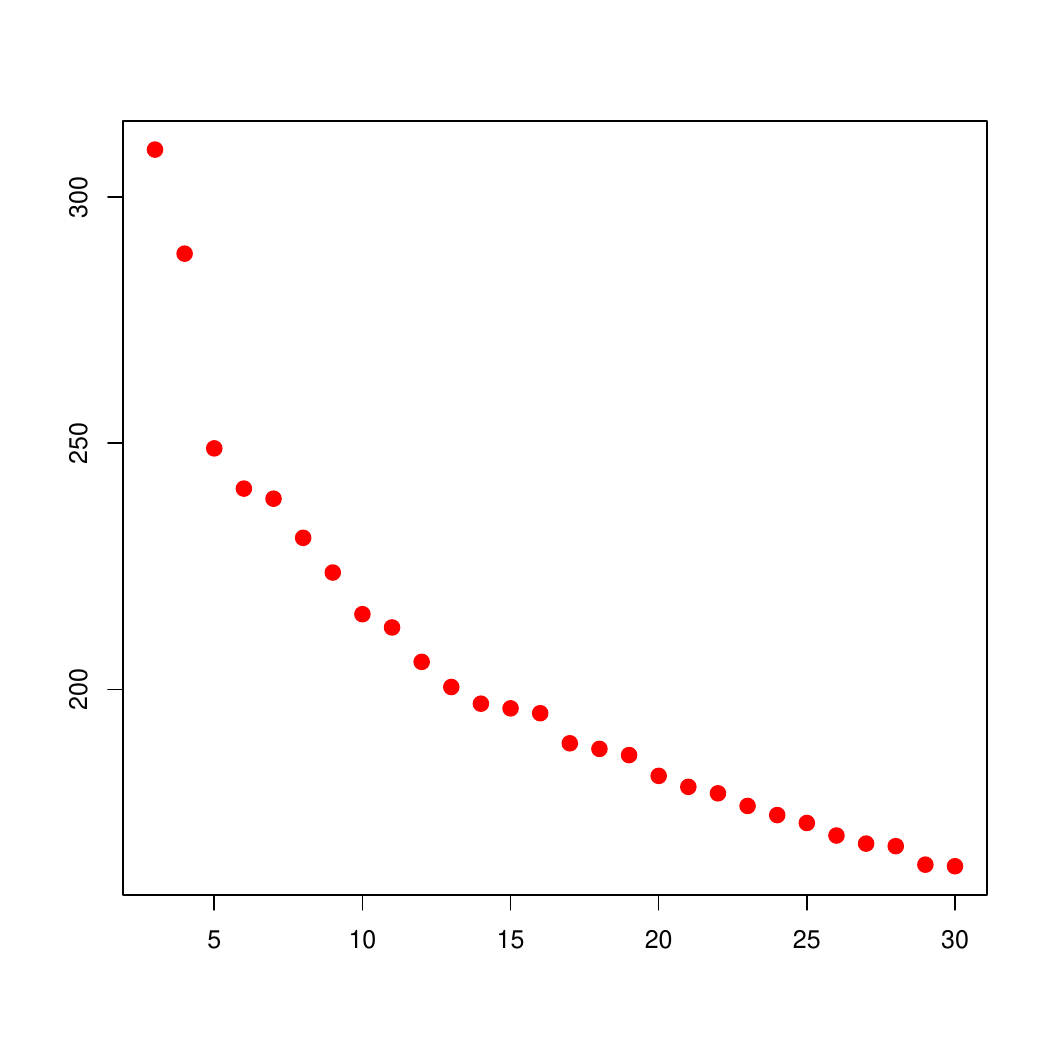}
\caption{Scree plot of the text corpus matrix $D$. Left: top 30 singular values. Right:  omitting first two singular values for a better visualization. } \label{fig:scree}
\end{figure}

First, we check the scree plot of the text corpus matrix $D$; see Figure~\ref{fig:scree}. The elbow points are 4 and 16. We thus consider the range of $4\leq K\leq 16$. 

Next, for each $4\leq K\leq 16$, we run topic-SCORE (Step 1 of TR-SCORE) to obtain $\hat{A}$ and then follow the approach in Section~\ref{sec:application} to find the most frequent anchor words for each topic. We use these anchor words to investigate the research areas covered by each discovered topic. 

For example, Table~\ref{tab:Topic-other-K-1} displays the 20 most frequent anchor words of each topic, based on the output of topic-SCORE for $K=4$ and $K=5$, respectively. We compare the two outputs and re-order the topics for $K=5$ so that the first 4 topics have a one-to-one correspondence to the topics for $K=4$. After checking the anchor words of the 5th topic for $K=5$ and using our knowledge of the field of statistics, we think this topic can be interpreted as ``Regression'' and is meaningful. We thus prefer $K=5$ to $K=4$. 

\spacingset{1.1}
\begin{table}[hbt!]
\centering
\caption{The 20 most frequent anchor words of each topic when $K=4$ (top) or $K=5$ (bottom). We have re-ordered topic labels so that the first 4 topics for $K=5$ have similar interpretations as the topics for $K=4$.} \label{tab:Topic-other-K-1}
\scalebox{0.85}{
\begin{tabular}{c P{16cm}}
Topic & Frequent anchor words \\
\hline
1 &  latin, doptim, block, nonregular, satur, resolut, orthogon, prime, array, cyclic, aoptim, neighbor, urn, divis, combinatori, extrapol, optim, search, incomplet, criteria\\
\hline
2 & agespecif, birth, frailti, longitudin, pollut, socioeconom, subjectspecif, timevari, wait, age, survivor, air, landmark, regist, missing, femal, day, tempor, geograph, nonignor\\
\hline
3 & noncompli, complianc,  antiretrovir, depress, physician, metaanalysi, particip, metaanalys, unmeasur, causal, timetoev, propens, prognost, intervent, therapi, chronic, symptom, coronari, patient, outcom\\
\hline
4 & cramervon, hotel, lagrang, goodnessoffit, distributionfre, onesampl, pvalu, cointegr, hypothes, onesid, chisquar, twosampl, stepdown, fdr, null, score, chi, pearson, diagnost, roc\\
\hline
\end{tabular}}

\vspace{.3cm}

\scalebox{0.85}{
\begin{tabular}{c P{16cm}}
Topic & Frequent anchor words \\
\hline
1 &  aoptim, doptim, latin, nonregular, twolevel, factori, aberr, twofactor, design, block, satur, prime, resolut, orthogon, cyclic, array, balanc, optim, column, divis 
\\
\hline
2 & hit, queue, semimarkov, traffic, statespac, forecast, evolutionari, shock, markov, repair, markovchain, renew, state, wind, mcmc, hidden, discretetim, segment, epidem, metropolishast\\
\hline
3 & noncompli, complianc, metaanalys, depress, causal, metaanalysi, unmeasur, outcom, prognost, particip, coronari, timetoev, surrog,  antiretrovir, dropout, physician, confound, smoke, elder, exposur\\
\hline
4 & cramervon, kolmogorovsmirnov, null, hotel, omnibus, test, goodnessoffit, lagrang, wald, hypothesi, wilcoxon, twosampl, distributionfre, onesampl, neyman, cointegr, pvalu, chisquar, ttest, permut\\
\hline
5 & regress, singleindex, ridg, backfit, explanatori, cook, lasso, spline, regressor, quantil, predictor, varyingcoeffici, curs, penalti, penal,  bspline, oracl, coeffici, tensor, variabl
\\
\hline
\end{tabular}}
\end{table}

\spacingset{1.475}

Similarly, we successively compare each pair of nested values of $K$. For each of $5\leq k\leq 11$, we find that increasing $K$ from $k-1$ to $k$ leads to the discovery of new topics that are meaningful. However, when we increase $K$ from $11$ to $12$, it is not the case. Table~\ref{tab:Topic-other-K-2} displays the 20 most frequent anchor words for each topic in the output of $K=12$. We use the anchor word list to match each topic with one of the 11 topics in the output of $K=11$ (see Figure~\ref{fig:anchor_words} of the main article). We find that 11 out of the 12 discovered topics can be matched to one of 11 topics in Figure~\ref{fig:anchor_words}. The 12th discovered topic (last row of Table~\ref{tab:Topic-other-K-2}) is not very meaningful to be listed as a new topic (the `anchor words' such as {\it rootn}, {\it longmemori}, {\it censorship} may be used by abstracts in different research areas of statistics). We thus prefer $K=11$ to $K=12$.
We also investigate $12< K\leq 16$ and find that these results are all less interpretable than that of $K=11$. We decide that $K=11$ is the most appropriate choice.

\spacingset{1.1}
\begin{table}[htb!]
\centering
\caption{The 20 most frequent anchor words of each topic when $K=12$. We have re-ordered topic labels so that the first 11 topics have similar interpretations as the topics for $K=11$ (see Figure~\ref{fig:anchor_words} of the main article).} \label{tab:Topic-other-K-2}
\scalebox{.85}{
\begin{tabular}{cl P{16cm}}
Topic & Manual label & Frequent anchor words\\
\hline
1 & Bayes & jeffrey, improp, frequentist, default, fuzzi, highthroughput, opinion, dirichlet, speci, probabilist, text, belief, pivot, protein, microarray, dna, census, genom, thousand, dissimilar" \\  
\hline
2 & Bio./Med. & epidemiolog, prospect, undertaken, alzheim, misclassif, environment, polymorph, ascertain, ecolog, retrospect, genomewid, smoke, matern, risk, conduct, suscept, coronari, occup, popul, missing   \\
\hline
3 & Clinic. & treatment, surrog, causal, propens, placebo, assign, unmeasur, effect, benefici, intervent, trial, imbal, subgroup, clinician, therapi, random, clinic, baselin, outcom, physician\\
\hline
4 & Exp.Design & aoptim, latin, design, twolevel, block, satur, nonregular, twofactor, factori, aberr, minimum, twophas, orthogon, fraction, resolut, experiment, balanc, multistag, doptim, divis\\
\hline
5 & Hypo.Test &  stepdown, familywis, fals, discoveri, bonferroni, twosid, cdf, reject, onesid, pvalu, conserv, realdata, hypothes, configur, competitor, microarray, nomin, favor, bootstrap, control \\
\hline
6 & Inference & confid, interv, width, shorter, biascorrect, edgeworth, coverag, squar, logarithm, rate, cap, underestim, mse, meansquar, pointwis, toler, upper, deconvolut, discontinu, slower \\
\hline
7 & Latent.Var. & proxi, instrument, forest, manifest, predictor, insur, latent, household, explanatori, exogen, sex, childhood, nonrespons, concomit, imput, variabl, interview, bernoulli, predict, enter \\
\hline
8& Mach.Learn.  & metropoli, boost, algorithm, particl, expectationmaxim, descent, faster, iter, svm, slow, updat, metropolishast, mcmc, step, sampler, path, noisi, gibb, heurist, nonsmooth \\
\hline
9& Math.Stats. & probab, expans, walk, nonneg, gumbel, mild, theorem, weak, ddimension, compact, equivari, trim, densiti, establish, element, omega, proof, press, stein, random \\
\hline
10 & Regression & regress, regressor, quantil, coeffici, smoother, band, least, calibr, shrink, linear, ordinari, logist, spline, backfit, scalar, influenti, equivari, leverag, leastsquar, error\\
\hline
11& Time Series & time, surviv, intervalcensor, gap, failur, multist, forecast, shock, censor, transplant, semimarkov, repair, periodogram, seri, occurr, event, declin, onset, drift, shortterm \\
\hline
12  & (unclear) & infinitedimension, nconsist, gamma, twoparamet, rootn, unknown, phi, inadmiss, nuisanc, mles, longmemori, weibul, threeparamet, ornsteinuhlenbeck, frailti, mestim, paramet, censorship, theta, semiparametr    \\
\hline
\end{tabular}}
\end{table}
\spacingset{1.475}

How to select $K$ in a topic model is a well-known challenging problem. To our best knowledge, there exists no method that works universally well. In theory, the singular values of $D$ (i.e., the scree plot) contain information of $K$ \citep{KaW2017}, but the scree plot of our data set is not informative enough for us to pin down the exact value of $K$ (see Figure~\ref{fig:scree}, where we only use the plot to determine a range of possible $K$). The perplexity \citep{BNJ2003} is a commonly used metric to assess the goodness-of-fit of a topic model. We may select $K$ by minimizing the perplexity, but this approach is known to be unstable  \citep{zhao2015heuristic}. It tends to select a very large $K$ on our data set, making the interpretation/labeling of topics difficult. Other ideas of estimating $K$ include the Bayesian approach which puts a prior on $K$ and computes the posterior, but it is unclear how to combine this idea with the topic-SCORE algorithm. We have tried many different approaches and found that the most satisfactory one is investigating the interpretability of discovered topics using our knowledge of the field, as described above.


\section{High-weight papers in each of the 11 topics} \label{subsec:purepapers}
In Section~\ref{sec:application}, we perform topic learning using the abstracts of $56,500$ papers and identify 11 topics. We propose a label for each topic using the topic loading vectors (see Figure~\ref{fig:anchor_words} of the main article). The short label is often insufficient to describe all the research topics that this topic covers. We further study each topic by investigating papers with high weights on this topic.

\spacingset{1.1}
\begin{table}[htb!]
\centering
\caption{For each of the 11 topics, the titles of the three papers that have the highest topic weight in that topic (last column: topic weight in that topic).} \label{tab:repre_papers}%
\scalebox{.8}{ \begin{tabular}{|c|P{16cm}|c|}   
\hline 
Topic & Title & Weight \\
\hline 
\multirow{3}[2]{*}{Bayes} & On Bartlett correction of empirical likelihood in the presence of nuisance parameters & 0.68\\
 & On the asymptotics of residuals in autoregressive moving average processes with one autoregressive unit root & 0.56\\
 & A note on universal admissibility of scale parameter estimators & 0.56\\
\hline 
\multirow{3}[2]{*}{Bio./Med.} & Analytic methods for 2-stage case-control studies and other stratified designs & 0.51\\
 & Reay and hope versus British Nuclear Fuels plc: issues faced when a research project formed the basis of litigation & 0.51\\
 & Statistical analysis in genetic studies of mental illnesses & 0.5\\
\hline 
\multirow{3}[2]{*}{Clinic.} & Estimating a multiplicative treatment effect under biased allocation & 0.61\\
 & Identifying and estimating net effects of treatments in sequential causal inference & 0.6\\
 & Advanced issues in the design and conduct of randomized clinical trials: the bigger the better? & 0.59\\
\hline 
\multirow{3}[2]{*}{Exp.Design} & Optimal block designs for triallel cross experiments & 0.91\\
 & An infinite family of non-embeddable quasi-residual designs & 0.89\\
 & Minimum aberration ($S^2$) $S^{n-k}$ designs & 0.89\\
\hline 
\multirow{3}[2]{*}{Hypo.Test} & A momentum-threshold autoregressive unit root test with increased power & 0.51\\
 & An unbiased test for the bioequivalence problem & 0.51\\
 & An example of a 2-sided wilcoxon signed rank test which is not unbiased & 0.5\\
\hline 
\multirow{3}[2]{*}{Inference} & Some inequalities in elementary special-functions with applications to nonparametric statistical-inference & 0.8\\
 & Increasing the confidence in Students t-interval & 0.79\\
 & Coverage-adjusted confidence intervals for a binomial proportion & 0.79\\
\hline 
\multirow{3}[2]{*}{Latent.Var.} & Variable selection in model-based clustering: a general variable role modeling & 0.76\\
 & The influence of variable selection - a Bayesian diagnostic perspective & 0.75\\
 & On Spitzer's formula for the moment of ladder variables & 0.66\\
\hline 
\multirow{3}[2]{*}{Mach.Learn.} & Java-ML: a machine learning library & 0.57\\
 & A genetic algorithm tutorial & 0.56\\
 & A gradient algorithm locally equivalent to the EM algorithm & 0.54\\
\hline 
\multirow{3}[2]{*}{Math.Stats.} & Comparison of level-crossing times for Markov and semi-Markov processes & 0.49\\
 & Estimation of conditional $L_1$-median from dependent observations & 0.48\\
 & Stochastic ordering of multivariate normal distributions & 0.47\\
\hline 
\multirow{3}[2]{*}{Regression} & Regression depth with censored and truncated data & 0.8\\
 & Continuum regression and ridge-regression & 0.73\\
 & The peculiar shrinkage properties of partial least squares regression & 0.67\\
\hline 
\multirow{3}[2]{*}{Time Series} & Some theoretical properties of the geometric and alpha-series processes & 0.83\\
 & Non-parametric estimation with doubly censored data & 0.75\\
 & Fitting semi-Markov models to interval-censored data with unknown initiation times & 0.75\\
\hline
\end{tabular}}%
\end{table}%
\spacingset{1.475}


For each $1\leq k\leq 11$, we sort the paper abstracts in the descending order of $\hat{w}_i(k)$.  Table~\ref{tab:repre_papers} shows the titles of the three abstracts with the largest $\hat{w}_i(k)$. The results are largely consistent with the proposed topic labels. Moreover, for each topic $k$, by reading the titles of the 300 papers with highest weights on this topic,  we come up with a list of suggested research topics umbrellaed by this topic. See Table~\ref{tab:11-topic} of the main article.

\section{The topic interests of 80 representative authors}\label{subsec:80authors}

In Section~\ref{sec:application},  we use the output of topic learning to define a {\it centered topic interest vector} $z_a\in\mathbb{R}^{11}$ for each author $a$. To recap, for each author $a$, let ${\cal N}_a\subset\{1,2,\ldots,n\}$ be the collection of papers published by this author  
in our data range, where each paper $i$ has an estimated 
topic weight vector $\hat{w}_i$ for its abstract. The centered topic interest vector $z_a$ is 
\[
z_a = \bar{w}_a - \bar{w},
\]
where $\bar{w}_a$ is the average of $\hat{w}_i$ over all abstracts in ${\cal N}_a$ and  $\bar{w}$ be the average of $\hat{w}_i$ over all ($n = 56,500$) abstracts. The entries of $z_a$ sum to $0$, and so it has both positive and negative entries. We are interested in positive entries of $z_a$: Author $a$ has greater-than-average weight on topic $k$ if $z_a(k) > 0$, for $1\leq k\leq 11$. See Figure~\ref{fig:tw_author_all} and details therein.

We now use $z_a$ to define the ``major topics" of author $a$ and show the results for 80 representative authors.  
Fix an author $a$. We call topic $k$ a ``major topic" of author $a$ if 
\[
z_a(k) > 50\% \times \max_{1 \leq \ell \leq 11} \{z_a(\ell)\}.
\]  
We may change $50\%$ to $(50\% \pm 5\%)$  but the results are similar.   

Table~\ref{tab:tw_80authors} presents the major topics of 80 authors with highest citations (ordered alphabetically).  We remark again that the short topic labels may not be accurate for all research areas each topic covers, and it is always useful to consult Table~\ref{tab:11-topic} of the main article.

\spacingset{1.02}
\setlength{\tabcolsep}{3pt}
\begin{table}[htb!]
\centering
\caption{The major topics for the 80 authors with highest citations (Topic $k$ is a major topic for author $a$ if $z_a(k)> 0.5\cdot \max_{1 \leq k \leq 11} \{z_a(k)\}$). The short topic labels such as `Mach.Learn.' may not be accurate to describe all research areas each topic covers, and it is always useful to consult Table~\ref{tab:11-topic} of the main article for the interpretation of each topic.  }  \label{tab:tw_80authors} 
\scalebox{.78}{
        \begin{tabular}{p{4.5cm} p{6cm}| p{4.5cm} p{6cm}}
        \toprule
        Name   & Major Topics  & Name   & Major Topics \\
        \midrule
Anderson, Per   &	Time Series	&  Little, Roderick &	Clinic., Latent.Var.\\
Azzalini, Adelchi  & Bayes, Mach.Learn.	& Louis, Thomas A. 	& Clinic.\\
Barndorff-nielsen, Ole & Math.Stats.	& Mammen, Enno	& Regression\\
Benjamini, Yoav& Hypo.Test, Inference	& Marron, James S. 	& Mach.Learn.\\
Berger, James 	& Bayes	& Mccullagh, Peter	& Bayes, Math.Stats.\\
Besag, Julian  &	Mach.Learn.	& Meng, Xiao-li	& Mach.Learn.\\
Best, Nicky  &	Bio./Med., Latent.Var., &	Molenberghs, Geert  & Clinic., Bio./Med.\\
	& Clinic., Mach.Learn. &	M\"uller, Hans-georg &	Regression\\
Bickel, Peter &	Math.Stats., Mach.Learn., & Owen, Art  &	Latent.Var., Math.Stats., \\
&	Regression	& &	Mach.Learn.\\
Breslow, Norman &	Regression	& Pepe, Margaret &	Bio./Med.\\
B\"uhlmann,, Peter &	Latent.Var.	& Prentice, Ross  &	Bio./Med.\\
Carlin, Bradley &	Mach.Learn., Clinic.	& Raftery, Adrian  &	Latent.Var., Mach.Learn.\\
Carroll, Raymond 	& Regression	& Rao, Jon N. K. 	& Inference, Regression\\
Clayton, David  &	Bio./Med.	& Rice, John	& Time Series, Mach.Learn.\\
Cook, Dennis	& Regression	& Roberts, Gareth & 	Mach.Learn.\\
Cox, David &	Latent.Var.&	Robins, James	& Clinic., Time Series\\
Dempster, Arthur P.	& Inference, Time Series	& Rosenbaum, Paul R. &	Clinic.\\
Dette, Holger	& Exp.Design	& Rotnitzky, Andrea	& Clinic.\\
Donoho, David	& Inference, Regression, 	& Rubin, Donald B. 	& Clinic.\\
	& Math.Stats., Mach.Learn.	& Ruppert, David	& Regression\\
Efron, Bradley &	Inference	& Silverman, Bernard W. 	& Mach.Learn., Regression\\
Fan, Jianqing	& Regression	& Smith, Adrian 	& Mach.Learn., Bayes\\
Fleming, Thomas R. 	& Clinic. &	Spiegelhalter, David 	& Clinic., Mach.Learn.\\
Friedman, Jerome 	& Mach.Learn., Latent.Var. &	Stone, Charles J.  &	Regression, Inference,\\
Gelfand, Alan	& Mach.Learn., Latent.Var.	&&	 Latent.Var.\\
Gill, Richard	& Math.Stats.	& Stute, Winfried 	& Regression, Hypo.Test,\\
Green, Peter	& Mach.Learn., Latent.Var.	&&	Math.Stats.\\
Hall, Peter 	& Inference	& Tibshirani, Robert	& Mach.Learn.\\
H\"ardle, Wolfgang  &	Regression	& Tsiatis, Anastasios  & Clinic.\\
Hastie, Trevor 	& Mach.Learn. &	Tsybakov, Alexandre  & Regression, Inference,\\
Ibrahim, Joseph & Clinic., Bayes, Bio./Med.	&&	 Math.Stats.\\
Johnstone, Iain	& Inference, Math.Stats.	& Wand, Matt P.  & Regression\\
Jones, M. C.  &	Math.Stats., Regression &	Ware, James &	Clinic., Latent.Var.\\
Kalbfleisch, John D.	& Time Series, Bio./Med. & Wasserman, Larry &  Inference, Mach.Learn.\\
Laird, Nan & Clinic.	& Wei, Lee-jen 	& Time Series\\
Lawless, Jerry 	&Time Series, Bio./Med.&	West, Mike  & Mach.Learn., Time Series\\
Li, Ker Chau 	& Regression &	Wu, Chien Fu	& Exp.Design\\
Li, Runze 	& Regression, Latent.Var.	& Ying, Zhiliang	& Time Series, Regression\\
Liang, Kung Yee 	& Bayes, Bio./Med. & Zeger, Scott 	& Clinic., Time Series\\
Lin, Danyu Y. 	& Time Series, Bio./Med.,	& Zhao, Lueping	& Bio./Med., Regression\\
&	 Regression	& Zhu, Lixing &	Regression, Hypo.Test\\
Lin, Xihong	& Clinic., Regression	  & Zou, Hui  & Latent.Var., Mach.Learn.,\\
Lindsay, Bruce 	& Bayes, Mach.Learn.	&	& Regression\\
Lipsitz, Stuart 	& Clinic., Regression, Bio./Med. &	&	\\
        \bottomrule
        \end{tabular}
}
\end{table}
\spacingset{1.475}

\section{Topic trends in 7 representative journals}

\begin{figure}[htb!]
\centering  
\includegraphics[width=.9\textwidth]{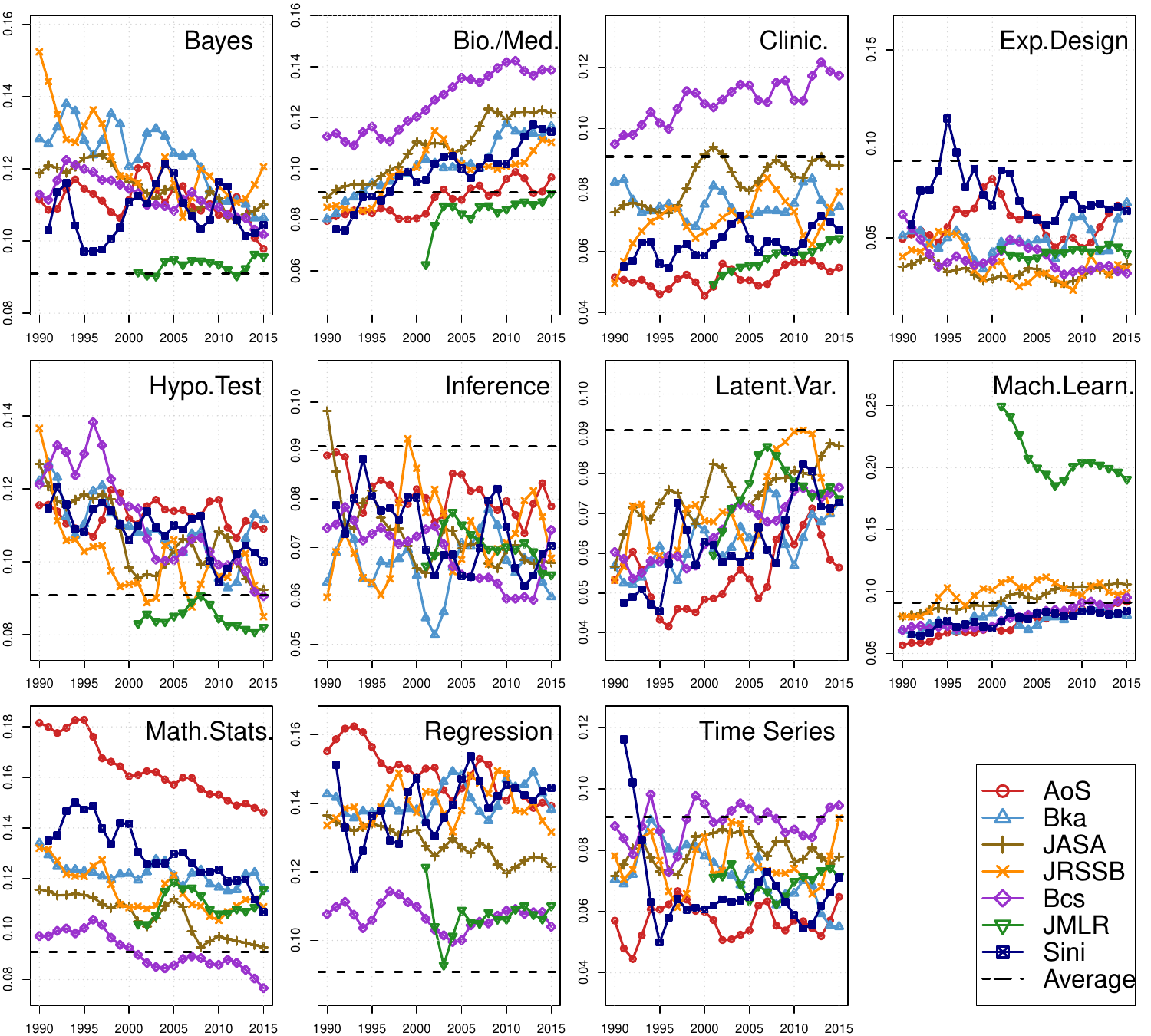}\\
\vspace{.5em}
\scalebox{.7}{
\begin{tabular}{ll | ll | ll | ll}
\hline
Topic  & Journal & Topic & Journal & Topic & Journal & Topic & Journal \\ 
\hline 
Bayes &  Bka & Exp.Design &  Sini & Latent.Var. &  JMLR (04-07)  &   Regression &  AoS (90-02)   \\ 
 &  JRSSB & Hypo.Test & Bcs (90-98) & &  JRSSB (08-12) &  Time Series &  Bcs\\
Bio./Med. &  Bcs  &  & AoS (02-15)  &  Mach.Learn. & JMLR &  \\ 
Clinic. &  Bcs   & Inference &  AoS &  Math.Stats.& AoS    \\
\hline \\ 
\end{tabular}
} 
\caption{The yearly average topic weights for $7$ selected journals during 1990--2015, and the friendliest journal (out of these 7 journals) for each topic.} \label{fig:tw_dynamic_journal}
\end{figure}

In Section~\ref{subsec:topic-TT}, we have selected a few journals and study how the evolution of the yearly average topic weights for each journal. 
Based on the journal ranking by the Stigler's model and PageRank (see Section~\ref{subsec:topic-JR}),  
we select the 7 journals with highest average ranks:  {\it AoS}, {\it Bka}, {\it JASA}, {\it JRSSB}, {\it Bcs}, {\it JMLR}, and {\it Sini}. For each journal, we obtain the yearly average topic weight (i.e., the average of $\hat{w}_i$ among papers published in this journal each year) and smooth the curves as before. The results are in Figure~\ref{fig:tw_dynamic_journal}. While we may plot the average weights of different topics in the same journal, we choose to plot the average weights of the same topic in different journals. In Figure~\ref{fig:tw_dynamic_journal}, each panel corresponds to a topic, and different curves in each panel represent different journals. 
 
%

\spacingset{1.1}
\spacingset{1.475}

\section{The cross-topic citation weights}
In Section~\ref{subsec:citation-graph}, we have introduced two definitions of the cross-topic-citation graph. In the first definition, for each $1\leq k\neq \ell\leq K$, there is a directed edge from topic $k$ to topic $\ell$ with weight $P_{k\ell}=N_{k\ell}/(\sum_{m=1}^K N_{km})$, where
\[
N_{k \ell} = \sum_{i, j =1 }^n \hat{w}_i(k) \hat{w}_j(\ell) C_{ij}. 
\]
In the second definition, for each $1\leq k\neq \ell\leq K$, there is a directed edge from topic $k$ to topic $\ell$ with weight $P^*_{k\ell}=N^*_{k\ell}/(\sum_{m=1}^K N^*_{km})$, where
\[
N^*_{k \ell} = \sum_{i, j =1 }^n \hat{w}^*_i(k) \hat{w}^*_j(\ell) C_{ij}, \qquad\mbox{with}\quad \hat{w}_i^*\in\{e_1,e_2,\ldots,e_K\}. 
\]
Here, $\hat{w}_i^*$ encodes the `dominant topic', i.e., $\hat{w}_i^*=e_k$ if and only if $k=\mathrm{argmax}\{1\leq m\leq K: \hat{w}_i(m)\}$. The two $11\times 11$ weighted adjacency matrices $P$ and $P^*$ are presented in Table~\ref{tb:cross-topic-citation} and Table~\ref{tb:cross-topic-citation-2}, respectively. 

\spacingset{1.1}
\begin{table}[tb!]
\centering

\caption{The cross-topic citation matrix $P^*$ (by dominant topics). The diagonal elements show the proportions of within-topic-citations. The off-diagonal elements that are $\geq 0.09$ are marked grey. This matrix is used to construct the graph in Figure~\ref{fig:TopicCitations} of the main article. } \label{tb:cross-topic-citation}
\scalebox{.72}{
\begin{tabular}{l | ccccccccccc}
\midrule
	&	Bayes	&	Bio./Med	&	Clinic.	&	Exp.Des	&	Hypo.Test	&	Inference	&	Latent.Var	&	Mach.Learn	&	Math.Stats	&	Regression	&	Time Seri.\\	
	\hline
Bayes	&	\underline{.230}	&	.057	&	.046	&	.013	&	.070	&	.056	&	.066	&	\colorbox{black!10}{.127}	&	\colorbox{black!10}{.130}	&	\colorbox{black!10}{.134}	&	.072\\
Bio./Med.	&	\colorbox{black!10}{.096}	&	\underline{.143}	&	\colorbox{black!10}{.099}	&	.029	&	.081	&	.048	&	.070	&	.081	&	.081	&	\colorbox{black!10}{.169}	&	\colorbox{black!10}{.101}\\
Clinic.	&	.076	&	\colorbox{black!10}{.090}	&	\underline{.339}	&	.050	&	.064	&	.034	&	.060	&	.061	&	.036	&	\colorbox{black!10}{.098}	&	\colorbox{black!10}{.091}\\
Exp.Design	&	.029	&	.049	&	.079	&	\underline{.562}	&	.056	&	.030	&	.034	&	.034	&	.039	&	.064	&	.024\\
Hypo.Test	&	.062	&	.048	&	.038	&	.019	&	\underline{.454}	&	.049	&	.038	&	.041	&	\colorbox{black!10}{.092}	&	\colorbox{black!10}{.112}	&	.048\\
Inference	&	.088	&	.054	&	.034	&	.026	&	\colorbox{black!10}{.103}	&	\underline{.242}	&	.064	&	.063	&	\colorbox{black!10}{.124}	&	\colorbox{black!10}{.148}	&	.054\\
Latent.Var.	&	\colorbox{black!10}{.092}	&	.053	&	.047	&	.014	&	.048	&	.046	&	\underline{.256}	&	\colorbox{black!10}{.116}	&	.079	&	\colorbox{black!10}{.203}	&	.046\\
Mach.Learn.	&	\colorbox{black!10}{.123}	&	.055	&	.039	&	.017	&	.048	&	.044	&	\colorbox{black!10}{.097}	&	\underline{.312}	&	.087	&	\colorbox{black!10}{.122}	&	.056\\
Math.Stats.	&	\colorbox{black!10}{.102}	&	.041	&	.018	&	.013	&	.068	&	.071	&	.077	&	.073	&	\underline{.347}	&	\colorbox{black!10}{.126}	&	.064\\
Regression	&	.073	&	.047	&	.030	&	.015	&	.055	&	.050	&	\colorbox{black!10}{.096}	&	.061	&	.087	&	\underline{.431}	&	.055\\
Time Series	&	.089	&	.072	&	.066	&	.013	&	.057	&	.045	&	.046	&	.076	&	\colorbox{black!10}{.090}	&	\colorbox{black!10}{.141}	&	\underline{.303}\\
\bottomrule
\end{tabular}
}

\vspace{.3cm}

\caption{The cross-topic citation matrix $P$ (by topic weights). The diagonal elements show the proportions of within-topic-citations. The off-diagonal elements that are $\geq 0.11$ are marked grey. This matrix is used to construct the graph in Figure~\ref{fig:TopicCitations-2}. } \label{tb:cross-topic-citation-2}
\scalebox{.72}{
\begin{tabular}{l | ccccccccccc}
\toprule
	&	Bayes	&	Bio./Med	&	Clinic.	&	Exp.Des	&	Hypo.Test	&	Inference	&	Latent.Var	&	Mach.Learn	&	Math.Stats	&	Regression	&	Time Seri.\\	
	\hline
Bayes	&	\underline{.125} 	&	.101 	&	.076	&	.031	&	.107	&	.072	&	.068	&	.092	&	\colorbox{black!10}{.123}	&	\colorbox{black!10}{.138}	&	.068	\\
Bio./Med.	&	\colorbox{black!10}{.113}	&	\underline{.108} 	&	.084	&	.034	&	.099	&	.072	&	.070	&	.090	&	\colorbox{black!10}{.117}	&	\colorbox{black!10}{.139}	&	.073	\\
Clinic.	&	\colorbox{black!10}{.111}	&	.107 	&  \underline{.108} 	&	.040	&	.096	&	.068	&	.071	&	.088	&	.109 	&	\colorbox{black!10}{.128}	&	.073	\\
Exp.Design	&	.088	&	.090	&	.080	&	\underline{.207}	&	.086	&	.063	&	.056	&	.075	&	.099	&	.108	&	.048	\\
Hypo.Test	&	\colorbox{black!10}{.119}	&	.099	&	.075	&	.033	&	\underline{.130}	&	.073	&	.063	&	.083	&	\colorbox{black!10}{.126}	&	\colorbox{black!10}{.138}	&	.063	\\
Inference	&	.108	&	.099	&	.073	&	.034	&	.101	&  \underline{.100}	&	.074	&	.091	&	\colorbox{black!10}{.118}	&	\colorbox{black!10}{.138}	&	.064	\\
Latent.Var.	&	\colorbox{black!10}{.110}	&	.100	&	.077	&	.031	&	.090	&	.077	&	\underline{.101}	&	.099	&	\colorbox{black!10}{.112}	&	\colorbox{black!10}{.138}	&	.064	\\
Mach.Learn.	&	\colorbox{black!10}{.115}	&	.102	&	.076	&	.034	&	.094	&	.076	&	.077	&	\underline{.110}	&	\colorbox{black!10}{.114}	&	 \colorbox{black!10}{.135}	&	.068	\\
Math.Stats.	&	\colorbox{black!10}{.116}	&	.100	&	.073	&	.033	&	.107	&	.073	&	.067	&	.086	&	\underline{.135}	&	\colorbox{black!10}{.144}	&	.065	\\
Regression	&	\colorbox{black!10}{.113}	&	.100	&	.072	&	.032	&	.100	&	.074	&	.070	&	.088	&	\colorbox{black!10}{.125}	&	\underline{.159}	&	.066	\\
Time Series	&	\colorbox{black!10}{.112}	&	.107	&	.083	&	.028	&	.092	&	.069	&	.067	&	.091	&	\colorbox{black!10}{.112}	&	\colorbox{black!10}{.132}	&	\underline{.106}	\\
\bottomrule
\end{tabular}
}

\end{table}
\spacingset{1.475}

\spacingset{1.1}
\begin{figure}[htb!]
\centering
\includegraphics[height=.5\textwidth, width=.6\textwidth]{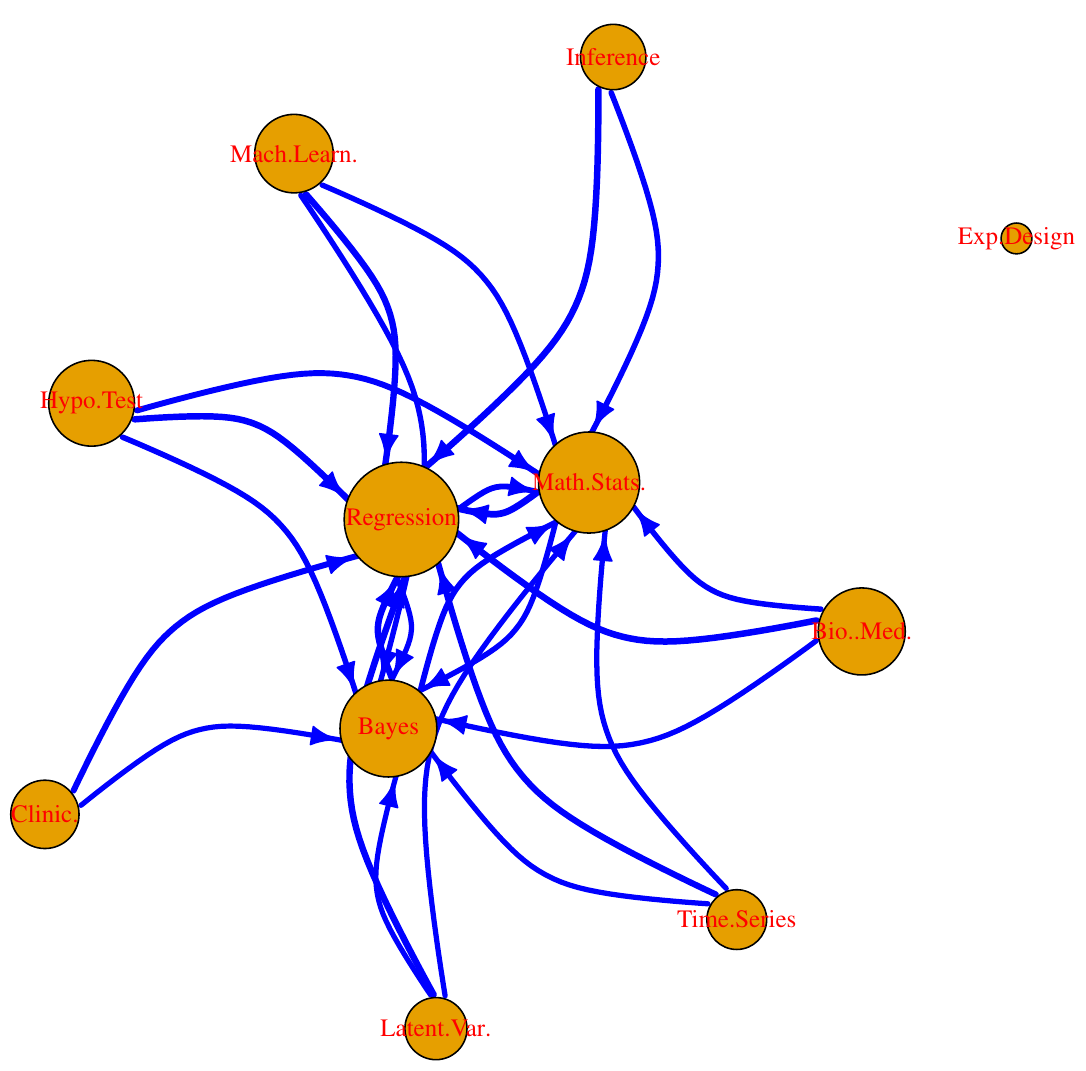} 
\caption{The cross-topic citation graph associated with $P^*$ (the width of edge if proportional to the weight of this edge; only edges with a weight $\geq 0.11$ are shown).} \label{fig:TopicCitations-2}
\end{figure}
\spacingset{1.475}

It is seen from Tables~\ref{tb:cross-topic-citation}-\ref{tb:cross-topic-citation-2} that distribution of elements in $P$ are more heavy tailed. As a result, if we apply the same threshold $P$ to $P^*$ to get two binary matrices, the one associated with $P$ is sparser and may be more interesting. For this reason, we choose to present the graph associated with $P$ (thresholded at $0.09$) in the main text; see Figure~\ref{fig:TopicCitations}. 
The graph associated with $P^*$ (thresholded at $0.11$) is shown in Figure~\ref{fig:TopicCitations-2}. 

The diagonal elements of $P$ and $P^*$ show the proportion of within-topic-citations for each topic. We observe that {\it Exp.Design}, {\it Hypo.Test}, {\it Math.Stats.} and {\it Regression} are the topics whose proportions of within-topic-citations are relatively high, and that {\it Bio./Med.}, {\it Inference} and {\it Latent.Var} are the topics whose proportions of within-topic-citations are relatively low.

\bibliographystyle{chicago}
\bibliography{reference} 

\end{document}